\documentclass[letterpaper,twocolumn,10pt]{article}
\usepackage{usenix}

\usepackage{tikz}
\usepackage{cite}
\usepackage{amsmath,amssymb,amsfonts}
\usepackage{algorithmic}
\usepackage{graphicx}
\usepackage{textcomp}
\usepackage{xcolor}
\def\BibTeX{{\rm B\kern-.05em{\sc i\kern-.025em b}\kern-.08em
    T\kern-.1667em\lower.7ex\hbox{E}\kern-.125emX}}
\usepackage{xspace}
\usepackage{makecell}
\usepackage{pifont}
\usepackage{multirow}
\usepackage{hyperref}
\usepackage{threeparttable}
\usepackage{booktabs}
\usepackage{url}

\newcommand{\ignore}[1]{}
\newcommand{\yes}{\tikz\draw[fill=black] (0,0) circle (.2em);}
\newcommand{\no}{\tikz\draw[fill=white] (0,0) circle (.2em);}
\usepackage{tcolorbox}
\newtcolorbox{findingbox}{
  left=1pt,
  right=1pt,
  top=1pt,
  bottom=1pt,
  boxrule=0.5pt,
  colback=white!0
}

\newcommand{\monet}{\textit{Monet}\xspace}

\usepackage{filecontents}

\begin{document}

\date{}

\title{\Large \bf \monet: Measuring the Ecosystem of Open-Source Text-to-Image Models Tailored for Harmful Services}

\author{
{\rm Zihao Wang}$^{1}$,
{\rm Jiacen Xu}$^{2}$,
{\rm Zilong Lin}$^{1}$\\
$^{1}$University of Missouri-Kansas City,
$^{2}$Microsoft
}

\maketitle

\begin{abstract}

The open-source text-to-image (T2I) ecosystem enables rapid model development and sharing, but also hosts models intentionally tailored for harmful services, which we call \textit{Monets}. Prior work has examined specific types of harmful T2I models on individual platforms, but a Monet does not exist in isolation. The broader Monet ecosystem, spanning model characteristics, cross-platform propagation, governance evasion, monetization, and downstream deployment, remains poorly understood.

In this study, we present the first systematic, ecosystem-level measurement of Monets. Grounded in the policies of real-world model hubs, we construct a taxonomy of ten harmful service categories and identify 23,947 Monets across eight major T2I model hubs, with the most popular exceeding 19 million downloads. While some developers employ anti-theft mechanisms against unauthorized re-uploading, Monets propagate across platforms at scale, with 40.76\% mirrored across hubs. Such propagation further enables governance evasion via cross-platform archiving, keeping 11.99\% of Monets accessible after bans on their original platforms, alongside other evasion strategies including keyword obfuscation and model-level safeguard circumvention. Monets also anchor coordinated commercial campaigns---one spanning 668 models with 914 completed commissions and another advertising gray-market account-farming service---and reach users through GitHub projects and inference APIs, raising downstream child safety concerns. These findings expose the limitations of platform-siloed defenses and highlight the need for cross-platform threat intelligence, coordinated governance, and technical safeguards.

\end{abstract}

\section{Introduction}~\label{sec:introduction}

Recent advances in generative AI, particularly text-to-image (T2I) generation, have significantly lowered the barrier to creating high-quality visual content. Beyond widespread adoption by end users, this progress has spawned a rapidly growing ecosystem that builds, fine-tunes, redistributes, and deploys T2I models for specialized purposes. Model-sharing hubs such as Civitai~\cite{Civitaipolicy}, LiblibAI~\cite{Liblibpolicy}, and Hugging Face~\cite{HFpolicy} further accelerate this ecosystem by offering infrastructure for hosting, discovering, and reusing open-source T2I models. While many of these models support benign creative applications, the same ecosystem also hosts models deliberately tailored for harmful services, explicitly prohibited or restricted by platform policies. We refer to such har\textbf{m}ful-service-tailored \textbf{o}pe\textbf{n}-sourc\textbf{e} \textbf{T}2I models as \textbf{\textit{Monets}}.

Prior work has touched on pieces of this problem: one study examined abusive models solely on Civitai, focusing on NSFW (Not Safe For Work) and deepfake models~\cite{wei2024exploring}, while others have proposed approaches for poisoning model hubs with harmful models~\cite{chen2026customization}. These efforts, however, focus on isolated corners of a much larger ecosystem. In practice, a Monet does not exist in isolation: it is \textit{developed} and \textit{hosted} on one platform, \textit{mirrored} across others, preserved through mechanisms that \textit{evade} platform moderation and model guardrails, \textit{monetized} via external storefronts and coordinated campaigns, and ultimately \textit{deployed} in downstream services. Yet little is known about this end-to-end ecosystem.

\noindent\textbf{Our work}. To address this gap, we present the first systematic, ecosystem-level measurement study of Monets. 
Specifically, we collected 23,947 Monets from eight major T2I model hubs---Civitai, CivArchive, LiblibAI, Hugging Face, ModelScope, SeaArt, Shakker, and TensorArt---spanning October 2022 to May 2026. To understand the Monet ecosystem, we investigate four research questions: \textbf{\textit{RQ1}}: what are the characteristics of Monets? \textbf{\textit{RQ2}}: how do Monets propagate across model hubs and evade governance mechanisms? \textbf{\textit{RQ3}}: how are Monets promoted and monetized across the ecosystem? \textbf{\textit{RQ4}}: how are Monets adopted and used in downstream applications and services?
Together, these questions paint an ecosystem-level picture of Monets, spanning their characteristics, cross-platform proliferation and persistence, commercial operations, and real-world downstream deployment.

\noindent\textbf{Our findings}.
Our study reveals that Monets not only represent a content-moderation challenge, but also constitute a sophisticated, multi-platform commercial ecosystem with resilient distribution and real-world harm.

\noindent\textit{ \underline{RQ1: Monet characteristics}}. Monets are widely distributed across all eight hubs, with Civitai and CivArchive alone hosting 79.33\% of collected models. Their popularity is substantial: 12 models such as \textit{WAI-NSFW-illustrious-SDXL}~\cite{highestdownloads} have accumulated over one million downloads. 75.74\% target sexual and nude content generation, followed by intellectual property infringement (37.85\%) and violence content (3.91\%) (see \S\ref{subsubsec:categorization_results}). The training data are equally concerning: developers not only rely on harmful datasets hosted on model hubs but also collect images directly from social media accounts of identifiable individuals (see \S\ref{subsubsec:training_data}). Some Monets insert sensitive terms in model metadata as an anti-theft method, exploiting strict platform moderation as a deterrent (see \S\ref{subsubsec:anti-theft}).

\noindent\textit{ \underline{RQ2: Propagation and governance evasion}}. Rather than being contained by platform moderation, Monets actively spread across platforms and evade governance. 40.76\% of collected Monets are mirrored across platforms by model hubs themselves, while 8.74\% carry developer-embedded model referral links maintaining cross-platform availability (see \S\ref{subsubsec:Platform-propagation}). Notably, CivArchive---a platform that publishes no content policy---has emerged as a centralized refuge, in which 11.99\% of the Monets had already been banned from their original platforms yet remain freely accessible (see \S\ref{subsubsec:archiving}). Also, we witnessed a sudden mass migration of Monets from Civitai to CivArchive on April 24, 2025, immediately following Civitai’s policy tightening the previous day (see \S\ref{subsubsec:distribution_growth}).

\noindent\textit{ \underline{RQ3: Promotion and monetization}}. Monets are not merely shared for free. They also anchor commercial operations (see \S\ref{subsec:monetization}). Developers embed external links in model metadata directing users to paid-access platforms, to sell premium model versions, custom training services, and created image sets. Beyond individual monetization, we uncover coordinated campaigns using the principle of guilt by association~\cite{wang2017gang}: one campaign of 668 Monets spans three platforms and operates a commission-based model-development service with 914 completed commissions; another uses deepfake and pornographic models as advertising channels for AI-assisted social-media account-farming services targeting TikTok and RedNote~\cite{Xiaohongshu}.

\noindent\textit{\underline{RQ4: Downstream deployment}}.
We identified 1,102 GitHub projects integrating Monets and two AI API providers providing APIs for 3,930 Monets. The GitHub projects range from research artifacts to NSFW chat. Several combine uncensored large language models (LLMs) with NSFW Monets to enable role-play interactions. In at least one case (see \S\ref{sec:deploy}), such interactions involve characters explicitly portrayed as minors, raising serious concerns about the potential generation of child sexual abuse material (CSAM).

\noindent\textbf{Contributions}. Our primary contributions are as follows:

\noindent$\bullet$\textit{ Ecosystem-level measurement}. 
We perform the first systematic measurement study of Monets from an ecosystem perspective. We characterize 23,947 Monets collected from eight major T2I model hubs, covering model characteristics, cross-platform propagation, governance evasion, monetization, and downstream deployment.

\noindent$\bullet$\textit{ Characterization of Monets and their resilience across platforms}. 
We characterize Monets across prevalence, technical foundations, and advertised harmful services, and empirically assess their capabilities of generating harmful content. 

We also show that, driven by both model hubs and developers, Monets propagate at scale across platforms and actively evade governance through cross-platform archiving, keyword obfuscation, and the circumvention of model-level safeguards.
 
\noindent$\bullet$\textit{ Exposure of Monet monetization and downstream abuse}. 
We reveal that Monets anchor organized monetization pipelines, including commission-based model-development and gray-market campaigns, orchestrated across platforms. We also show Monets reach real-world deployments through GitHub projects and hosted inference APIs, with downstream deployment raising potential child sexual abuse material risks.

\section{Background and Related Work}~\label{sec:background}

\subsection{Text-to-Image Models and Their Open-Source Ecosystem}\label{subsec:t2i_models}

Recent years have witnessed rapid progress in T2I generation, largely driven by advances in generative modeling.
Early text-conditioned image generation relied on GAN-based and autoregressive Transformer-based models~\cite{reed2016generative,zhang2017stackgan,bie2024renaissance}, which were often constrained by limited image fidelity. 
Modern open-source T2I models are predominantly built on two generative paradigms~\cite{holderrieth2025introduction}: denoising diffusion and flow matching, both of which generate images by progressively transforming noise into structured outputs.
The release of Stable Diffusion~\cite{rombach2022high}, a representative model of the denoising diffusion paradigm, marked a pivotal moment for the open-source T2I ecosystem by enabling high-quality image generation with relatively modest computational requirements. 
Many popular T2I base models follow this denoising diffusion paradigm, including SDXL~\cite{podell2024sdxl} and derivatives built upon it, such as Illustrious~\cite{park2024illustrious} and NoobAI~\cite{noobai}.
Flow matching, a closely related generative paradigm, was later adopted by open-source large-scale T2I models such as Stable Diffusion 3.5 and FLUX.1. 
In this study, we summarize the distribution of base-model generative paradigms and families among the collected Monets in \S\ref{subsubsec:basemodel}.

The open release of these T2I models has fostered a large community of developers who create fine-tuned variants, Low-Rank Adaptation (LoRA) modules~\cite{hu2022lora}, and task-specific checkpoints. As a result, model hubs such as Civitai, CivArchive, and Hugging Face have emerged as major platforms for sharing T2I models. Unlike conventional model hubs that host model files directly, CivArchive operates as a cross-platform model aggregator, providing model metadata and referral links to model files hosted on other platforms~\cite{CivArchivepolicy}.

\subsection{Misuse of Text-to-Image Models}

To trigger T2I models to create harmful content, various attack approaches have been proposed, such as jailbreaking~\cite{11023413,yang2024sneakyprompt}, plugin poisoning~\cite{chen2026customization} and concept restoration~\cite{zhang2024generate}.

In addition, T2I models have been widely misused in harmful applications such as image-based sexual abuse~\cite{hao2024harm}, deepfake generation~\cite{abdullah2024analysis}, hateful content~\cite{qu2023unsafe}, and intellectual property infringement~\cite{carlini2023extracting}. The prevalence of open-source T2I models has further lowered technical barriers, making such capabilities increasingly accessible and scalable.

More recently, Wei et al.~\cite{wei2024exploring} conducted an empirical analysis of Civitai, investigating abusive generative AI models, especially deepfake and NSFW-oriented models, and the produced images hosted on Civitai. However, existing studies have largely focused on specific categories of harmful models themselves and individual platforms. Consequently, the broader ecosystem of harmful T2I models in practice remains underexplored.

Our work fills this gap via a systematic ecosystem-level study of T2I models tailored for harmful services across major open-source T2I model hubs. Grounded in the explicit usage policies of these hubs, we build a taxonomy of prohibited harmful services and characterize the identified Monets across multiple dimensions: model characteristics, propagation, governance evasion, monetization, and real-world deployment.

\subsection{Threat Model}\label{subsec:threat_model}

\noindent\textbf{Threat scenario}. 

We consider an adversarial ecosystem in which developers tailor open-source T2I models capable of generating harmful content (i.e., Monets) and publish them on open-source T2I model hubs in violation of platform policies~\cite{Civitaipolicy,Shakkerpolicy,Liblibpolicy,Tensorpolicy,ModelScopepolicy,HFpolicy}. 

These Monets explicitly advertise their harmful purposes, while developers distribute them across multiple hubs, circumvent platform governance mechanisms, and monetize them through external services and communities. 
Downstream actors, including users and API providers, may further adopt these Monets to provide on-demand image-generation services. 

Within this ecosystem, the mechanisms of propagation and archival across model hubs allow Monets to remain accessible even after removal from individual platforms.

\noindent\textbf{Research scope}. 
Our study focuses exclusively on publicly accessible open-source T2I models explicitly advertised as tailored for harmful services and the ecosystem surrounding them. We exclude models exclusively designed for image-to-image or text-to-video generation, as well as closed-source T2I models. Although prior work has assessed the harmful capabilities of general-purpose T2I models~\cite{qu2023unsafe}, our objective is fundamentally different: instead of studying harmful capabilities of general-purpose T2I models, we explore the ecosystem of T2I models intentionally tailored for harmful services.

The harmful services considered in this work are derived from the usage policies of the major open-source T2I model hubs and organized into ten categories (see \S\ref{subsec:taxonomy}).

\section{Data Collection}~\label{sec:data_collection} 

In this section, we present our pipeline for discovering open-source T2I models explicitly advertised as tailored for harmful services across T2I model hubs.

\begin{figure}[t]
  \centering
  \includegraphics[width=1.0\linewidth]{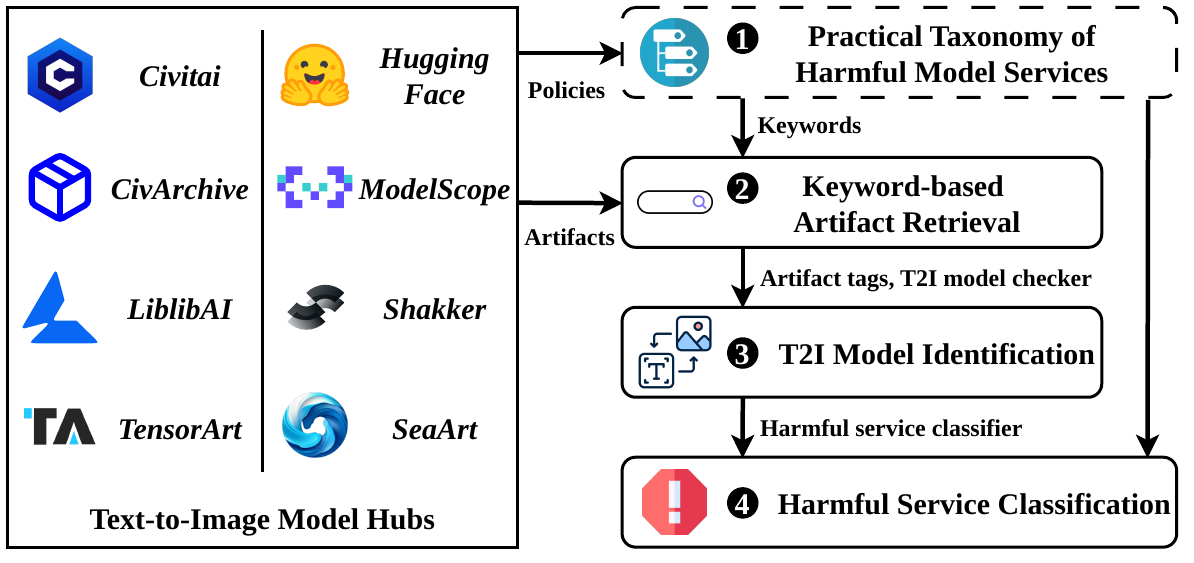}
  \caption{Workflow of discovering Monets from open-source T2I model hubs.}
  \label{fig:pipeline}
\end{figure}

\subsection{Overview}~\label{subsec:overview}

To systematically study the Monet ecosystem, we collected Monets from eight major hubs hosting T2I models---Civitai, CivArchive, LiblibAI, Hugging Face, ModelScope, Shakker, TensorArt, and SeaArt---following prior work~\cite{wei2024exploring,zeng2024intenttuner,civitai-mirror-list}.

Figure~\ref{fig:pipeline} illustrates an overview of the Monet discovery workflow. It includes practical taxonomy construction (\raisebox{-0.9pt}{\ding{182}}) and a discovery and classification pipeline consisting of three stages: keyword-based artifact retrieval (\raisebox{-0.9pt}{\ding{183}}), 
T2I model identification (\raisebox{-0.9pt}{\ding{184}}), and 
harmful service classification (\raisebox{-0.9pt}{\ding{185}}).

We begin by constructing a practical taxonomy of harmful model services by systematically analyzing prohibited or restricted services explicitly defined in the usage policies of real-world model hubs. The resulting taxonomy is summarized in Table~\ref{tab:taxonomy}.
Then, based on this taxonomy, we derive harmful-service-related keywords and retrieve candidate models from these hubs. 
Finally, we identify T2I models from retrieved results and classify them into harmful service categories based on their advertised services in model metadata, which are often documented in the form of model cards~\cite{mitchell2019model}, including titles, tags, descriptions, and demo content.

\subsection{Methodology}

\noindent\textbf{Constructing practical taxonomy of harmful model services}.\label{subsec:taxonomy}
Defining ``harmful'' services is challenging, as legal regulations and social norms vary significantly across jurisdictions. For example, pornographic content may be legally permissible in some countries but strictly prohibited in others. 

Rather than attempting to establish a universal definition of harmfulness, we adopt an ecosystem-grounded definition  
and define \textit{harmful model services} as services that are explicitly prohibited or restricted by major model hubs within the open-source T2I ecosystem. To operationalize this definition, we systematically analyzed the governance policies of seven major T2I model hubs (i.e., Civitai, LiblibAI, Hugging Face, ModelScope, Shakker, TensorArt, SeaArt)~\cite{Civitaipolicy,Shakkerpolicy,Liblibpolicy,Tensorpolicy,ModelScopepolicy,HFpolicy,seaartolicy}, representing the dominant norms and restrictions of this ecosystem. 
We aggregate the services prohibited or restricted by these policies and treat their union as the harmful services recognized by the open-source T2I ecosystem.
Note that CivArchive publishes no usage policy or content guidelines, which reflects its positioning as an unrestricted archive (see \S\ref{subsubsec:archiving})~\cite{CivArchivepolicy}.

To characterize such services, two experts independently annotated these policy provisions from seven major hubs. The annotators grouped semantically related services into unified categories, achieving high inter-annotator agreement (Cohen's $\kappa = 0.90$). Disagreements were resolved via discussion.
In this way, we identified 10 categories of harmful model services commonly recognized across the open-source T2I ecosystem in the real world, forming our \textit{practical taxonomy of harmful model services}, as listed in Table~\ref{tab:taxonomy}.

\noindent\textbf{Discovering and classifying Monets}.\label{subsec:discovering}
As outlined in \S\ref{sec:introduction}, our study focuses on T2I models intentionally tailored to provide harmful services, rather than general-purpose T2I models. Our goal is thus to identify models explicitly advertised for harmful purposes by developers, via a three-stage discovery pipeline:
(1) keyword-based artifact retrieval, 
(2) T2I model identification, and 
(3) harmful service classification.
Given that model versions may contain distinct model cards and files, we count different versions and cross-hub copies separately.

\noindent$\bullet$\textit{ Keyword-based artifact retrieval}.
Based on the practical taxonomy, we manually derived 99 keywords by extracting representative terms from hubs' policies for each harmful service category. Examples include ``NSFW,'' ``nudify,'' ``weapon,'' and ``suicide.''
We selected 10 keywords per category, except Category H, for which only nine were available due to limited relevant policy provisions.
We used these keywords to query the search engines of model hubs and collected the returned artifacts. 
Note that as our study focuses on the publicly advertised Monet ecosystem rather than exhaustively enumerating all harmful T2I models (\S\ref{subsec:threat_model}), 
keyword-based retrieval aligns with our scope by targeting models that expose explicit harmful-service signals in searchable metadata.

\noindent$\bullet$\textit{ T2I model identification}.
Because model hubs often host diverse artifacts beyond T2I models, including LLMs, embeddings, datasets, and repositories, we further identified T2I models from retrieved results. 
Specifically, following prior work~\cite{wei2024exploring}, we extracted checkpoint models (i.e., full model weights) and LoRA models (i.e., fine-tuned adapters) as \textit{candidate models} based on artifact types tagged in metadata, yielding 42,906 candidate models.
We then applied a T2I model checker assisted by a multimodal LLM (MLLM) to determine whether each candidate model in the form of checkpoint or LoRA supports T2I generation, by analyzing model metadata including titles, tags, descriptions, and demo content.
The checker uses a 3-point confidence rubric\footnote{Higher confidence scores indicate stronger evidence that the candidate model supports T2I generation according to its metadata.}, treating a model as T2I if its metadata either (1) explicitly states support for T2I generation or (2) showcases demo images created from text prompts. 
We retained artifacts that received the highest confidence score as T2I models.
In this way, we gathered a total of 28,948 T2I models.

\noindent$\bullet$\textit{ Harmful service classification}.\label{subsubsec:malicious_service_classification}
Finally, we classified the collected T2I models into 18 groups---17 harmful service subcategories summarized in Table~\ref{tab:taxonomy} and one harmless group.  
As the de facto standard for documenting AI models, model cards~\cite{mitchell2019model} describe models' purposes and have been used for such analysis in prior studies~\cite{lin2025consiglieres,puhlfurss2025model}. Thus, we use an MLLM-assisted classifier to analyze model metadata within model cards.
Since a single model may support multiple harmful services, we adopt a multi-label setting, i.e., the classifier assigns subcategories using a 4-point confidence rubric\footnote{Higher confidence scores indicate stronger evidence that the T2I model offers a service related to a harmful subcategory based on its metadata.}, retaining all subcategories that receive the highest confidence score. Models assigned at least one harmful service subcategory are identified as \textit{Monets}, 
while those not assigned any harmful service subcategory with the highest confidence score are placed in the harmless group.
In this way, we identified a total of 23,947 Monets from eight major T2I model hubs\footnote{The data was updated as of May 31, 2026}, including 10,678 in Civitai, 8,318 in CivArchive, 1,708 in LiblibAI, 621 in Hugging Face, 20 in ModelScope, 806 in SeaArt, 1,015 in Shakker, and 781 in TensorArt.

\begin{table*}[t!]
\centering
\scriptsize
\setlength{\tabcolsep}{2.pt}
\caption{Practical taxonomy of harmful model services summarized from T2I model hubs.}
\label{tab:taxonomy}
\begin{threeparttable}
\begin{tabular}{llccccccccc}

\toprule
\multirow{2}{*}{\textbf{Category}} &                        
\multirow{2}{*}{\textbf{Subcategory}} &
\multirow{2}{*}{\textbf{\# Monets}\tnote{*}} &
\multicolumn{8}{c}{\textbf{Model hubs\tnote{**}}} \\
\cmidrule(lr){4-11} & & &
{Civitai} &
{\makecell{Hugging Face}} &
{Shakker} &
{LiblibAI} &
{\makecell{TensorArt}} &
{\makecell{ModelScope}} &
{SeaArt} &
{\makecell{CivArchive}}\\

\midrule
\multirow{5}{*}{Sex and Nudity}
& A1. Minors in sexual contexts
&  1,105
& \yes & \yes & \no & \yes & \no & \yes & \yes & -- \\

& A2. Coercive/illegal sexual acts
& 949
& \yes & \no & \no & \yes & \no & \yes & \yes & -- \\

& A3. Explicit organs/fluids/fetish
& 9,623
& \yes & \no & \no & \yes & \no & \yes & \yes & -- \\

& A4. Sexualized posing/angles/props
& 17,499
& \no & \no & \no & \yes & \no & \yes & \yes & -- \\

& A5. Realistic-style sexual content
& 5,934
& \no & \no & \no & \yes & \no & \yes & \yes & -- \\

\midrule

Real-person likeness/deepfakes & B1. Real-person likeness/deepfakes 
& 935
& \yes & \yes & \yes & \yes & \yes & \yes & \yes & -- \\

\midrule

\multirow{3}{*}{Violence, gore, and horror}
& C1. Death/dismemberment/injuries
& 128
& \yes & \yes & \no & \yes & \no & \no & \yes & -- \\

& C2. Cruel violence and abuse
& 731
& \yes & \yes & \no & \yes & \no & \no & \yes & -- \\

& C3. Horror and terror
& 197
& \no & \no & \no & \yes & \no & \no & \no & -- \\

\midrule

Hate, harassment, and extremism & D1. Hate, harassment, and extremism
& 1
& \yes & \yes & \yes & \yes & \yes & \yes & \yes & -- \\

\midrule

\multirow{2}{*}{Illegal and regulated activities}
& E1. Illegal drugs and regulated goods
& 14
& \no & \yes & \yes & \yes & \yes & \no & \yes & -- \\

& E2. Intimidating criminal conduct 
& 46
& \yes & \yes & \yes & \yes & \yes & \no & \yes & -- \\

\midrule

Self-harm and dangerous behaviors & F1. Self-harm and dangerous behaviors
& 18
& \yes & \no & \no & \yes & \no & \no & \yes & -- \\

\midrule

Political misinformation & G1. Political misinformation
& 20
& \no & \no & \yes & \yes & \yes & \yes & \yes & -- \\

\midrule

Health and medical information & H1. Health and medical information
& 4
& \no & \no & \no & \no & \no & \no & \yes & -- \\

\midrule

Intellectual property infringement & I1. Intellectual property infringement
&  9,065
& \yes & \yes & \yes & \yes & \yes & \yes & \yes & -- \\

\midrule

Spam and deception & J1. Spam and deception
& 58
& \no & \no & \yes & \yes & \yes & \no & \yes & -- \\

\bottomrule
\end{tabular}

\begin{tablenotes}              
\item[*] Each identified Monet belongs to one or more subcategories. Subcategories within the same category are not mutually exclusive. 
\item[**] \yes\, indicates that the corresponding hub's policy prohibits or restricts the subcategory; \no\, otherwise. The taxonomy aggregates such policy-defined services across the seven hubs.
\end{tablenotes}

\end{threeparttable}
\end{table*}

\subsection{Evaluation}\label{subsec:evaluation}

\noindent\textbf{Implementation.}
In this study, we employed GPT-5~\cite{gpt5} as the MLLM backbone for both the T2I model checker and the harmful service classifier.

\noindent\textbf{Validation}.
To evaluate the effectiveness of our pipeline, we conducted post-hoc human validation for both T2I model checking and harmful service classification. Two experts with prior experience in AI ecosystems independently annotated the sampled artifacts. They resolved disagreements via discussion, forming the ground truth for evaluation. The annotation criteria for these tasks were aligned with the system prompts of the T2I model checker and harmful service classifier.

\noindent$\bullet$\textit{ T2I model checking}. We randomly sampled 500 models from 42,906 candidate models. Two human annotators achieved an inter-annotator agreement of Cohen’s $\kappa = 0.88$. Compared to the ground truth, the T2I model checker achieved an accuracy of 92.00\%, with a precision of 100.00\% and a recall of 90.15\%.

\noindent$\bullet$\textit{ Harmful service classification}. 
We randomly sampled 500 models from 28,948 T2I models. Since harmful service classification adopts a multi-label setting, we reported raw agreement rates. 
Human annotators achieved a raw inter-annotator agreement rate of 93.65\%. Against the ground truth, the harmful service classifier achieved an accuracy of 90.00\%, with a precision of 
88.19\% and a recall of 93.72\%.

\section{Landscape of Monets}\label{sec:measurement}

\subsection{Scope and Magnitude}

\begin{figure*}[t!]
    \centering
    \includegraphics[width=0.9\linewidth]{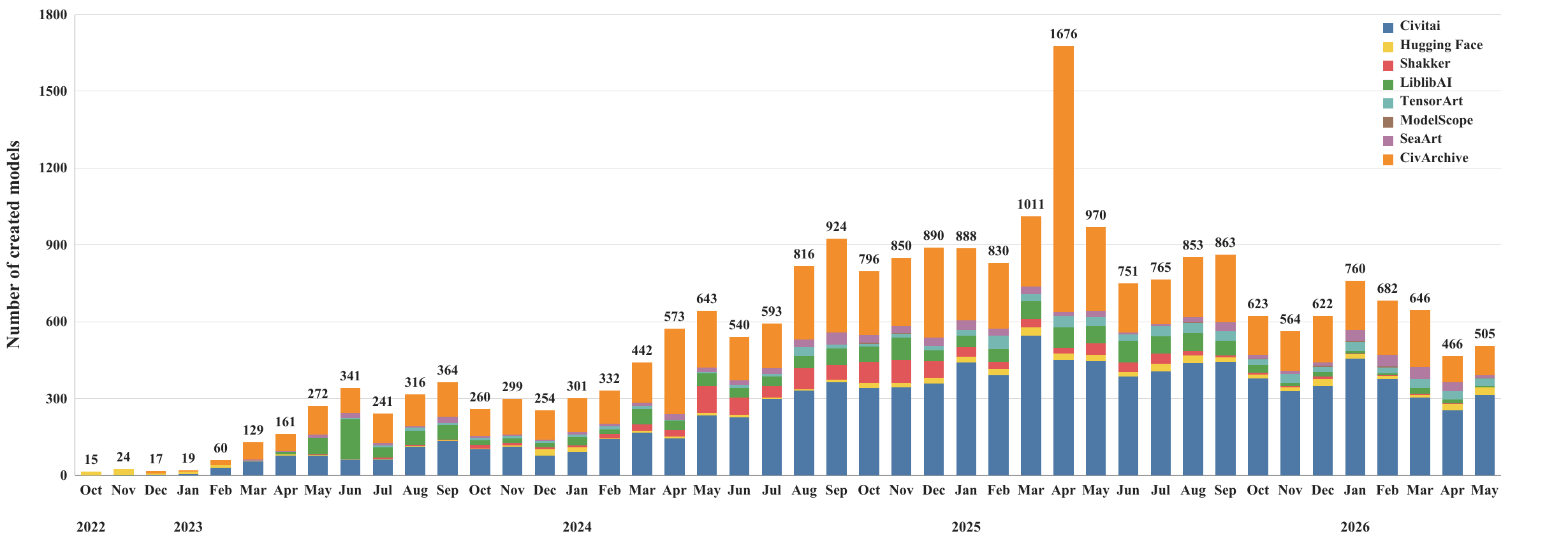}
    \caption{Temporal growth of Monets newly created on eight T2I model hubs.}
    \label{fig:monets-created}
\end{figure*}

\noindent\textbf{Distribution and growth}.\label{subsubsec:distribution_growth}
As mentioned in \S\ref{sec:data_collection},
we identified 23,947 Monets across eight major T2I model hubs. As shown in Fig.~\ref{fig:monets-created}, the earliest appeared on Hugging Face in October 2022, including \textit{sd-naruto-diffusers}~\cite{earlymodel2}, \textit{sd-pokemon-diffusers}~\cite{earlymodel3}, and \textit{Zack3D\_Kinky-v1}~\cite{earlymodel1}, tailored for intellectual-property characters and fetish-related content.

Among the surveyed hubs, Civitai hosts the largest number of Monets (10,678), followed by CivArchive (8,318) and LiblibAI (1,708), suggesting that Civitai serves as the primary hub for Monet hosting. CivArchive has also emerged as a major hub, due to its aggregation-oriented design and the absence of published content restrictions (see \S\ref{subsec:propagation} and \S\ref{subsubsec:archiving}).

As shown in Fig.~\ref{fig:monets-created}, the largest spike in Monet creation occurred in April 2025, with 61.87\% of Monets created in that month hosted on CivArchive. Based on their metadata, 78.91\% of them were uploaded to CivArchive on April 24, 2025, originating from Civitai. This indicates a sudden, large-scale migration of Monets from Civitai to CivArchive. 
Notably, this migration closely coincides with Civitai's policy update on April 23, 2025, which introduced stricter content-safety requirements~\cite{civitai_policy_adjust}, suggesting that policy changes may have influenced developers' platform-selection decisions and contributed to the observed migration of Monets.
Meanwhile, Civitai's policy change had little impact on the number of newly-created Civitai Monets, decreasing slightly from 451 in April 2025 to 446 in May.
Details of Monets' cross-platform propagation and archiving are discussed in \S\ref{subsec:propagation} and \S\ref{subsubsec:archiving}.

\begin{findingbox}
\textit{\textbf{Takeaway 1.}} 
A rapid migration of Monets to CivArchive, a less-moderated platform, immediately followed Civitai's policy update on April 23, 2025, highlighting how cross-platform migration can coincide with stricter platform governance while allowing harmful models to remain accessible elsewhere.

\end{findingbox}

\noindent\textbf{Usage and popularity}.
We collected downloads and ``like'' counts from each Monet's metadata. 
Monets average 8,801.09 downloads and 143.36 “likes” each.
As shown in Table~\ref{tab:downloads}, CivArchive exhibits the highest average downloads, while Civitai leads in average ``likes,'' suggesting that both platforms play an important role in the dissemination and user engagement of Monets within the open-source T2I ecosystem.
At the model level, \textit{WAI-NSFW-illustrious-SDXL}~\cite{highestdownloads} and \textit{Murky's - After Sex Lying LoRA}~\cite{highestlikes}, both targeting sexual and nude content, have the largest numbers of downloads (19 million) and ``likes'' (14.9K), respectively, reflecting strong user demand for such content within the ecosystem.

\begin{table}[t!]
\centering
\scriptsize
\begin{threeparttable}
\caption{Average downloads and ``likes'' of Monets on T2I model hubs.
}
\label{tab:downloads}
\begin{tabular}{l|lr|lr}

\toprule
\textbf{Rank} & \multicolumn{1}{c}{\textbf{Model hub}} & \multicolumn{1}{c}{\textbf{Ave. downloads}\tnote{*}} & \multicolumn{1}{c}{\textbf{Model hub}} & \multicolumn{1}{c}{\textbf{Ave. ``likes''}} \\

\midrule
1 & CivArchive & 23,023.41 & Civitai & 201.85\\
2 & Civitai & 1,549.27 & SeaArt & 31.64\\
3 & Hugging Face & 699.02 & TensorArt & 29.81\\
4 & ModelScope & 580.70 & Hugging Face & 14.76\\
5 & SeaArt & 172.20 & LiblibAI & 12.94\\
6 & LiblibAI & 116.87 & ModelScope & 8.20\\
7 & Shakker & 42.75 & Shakker & 4.91\\
8 & TensorArt & 2.79 & CivArchive & N/A\tnote{**}\\

\bottomrule
\end{tabular}
\begin{tablenotes}   
\item[*] 184 Monets on LiblibAI and 32 on Shakker do not disclose their downloads.
\item[**] CivArchive does not record the ``likes'' count of each model.
\end{tablenotes}
\end{threeparttable}
\end{table}

\noindent\textbf{Authorship}.
Based on the metadata of the collected Monets, we identified 7,372 developers with unique identifiers in the eight T2I model hubs, with Civitai accounting for the largest share (37.56\%). 
Monet contribution is highly uneven across developers: 14 are each responsible for over 100 models.

\noindent\textbf{Anti-theft mechanism}.\label{subsubsec:anti-theft} 
We found that four developers embed 107 politically sensitive Chinese phrases in the model cards of 12 models hosted on Civitai and CivArchive to defend against model theft and scraping. This long but incoherent set of phrases involves Chinese political incidents, politicians, and geopolitics and is unrelated to the model itself. 
The developers explicitly describe these phrases as an anti-theft mechanism in model cards. The distinctive text allows them to trace model re-uploads (see \S\ref{subsec:propagation}), particularly unauthorized ones, through online searches, as exemplified by \textit{Female POV}~\cite{female-pov} and its re-upload on PixAI~\cite{female-pov2}. Also, when propagated to platforms with strict content moderation, the embedded political terms may trigger automated review or removal of these copies.

\begin{findingbox}
\textit{\textbf{Takeaway 2.}} 
Some Monet developers embed a long but incoherent paragraph of sensitive keywords in model cards as an anti-theft mechanism, exploiting platform moderation systems to deter or penalize model re-uploading.
\end{findingbox}

\noindent\textbf{Textual triggers}.
The textual trigger is a sequence of one or more specific tokens whose presence in an input prompt activates a learned style, character, or concept of a model~\cite{mastering-trigger-words,struppek2023rickrolling}.
Among the Monets analyzed, 18,483 (77.19\%) disclose identifiable textual triggers, either through structured metadata fields or model descriptions. Specifically, 17,950 Monets provide triggers in structured fields, while an additional 533 explicitly specify triggers in their descriptions.

Among the Monets with textual triggers, 13,141 (71.10\%) specify multiple triggers, with one model \textit{Sui-Feng}~\cite{Sui-Feng} providing as many as 499. In total, we collected 202,617 trigger instances, corresponding to 45,767 unique triggers and an average of 11 triggers per Monet. The most prevalent triggers are ``1girl'' (3,020 Monets), ``long hair'' (2,152), and ``solo'' (1,387).
Triggers also vary substantially in complexity: 43.12\% (87,362) consist of a single token, while the remainder contain multiple tokens. The longest trigger contains 94 tokens and is associated with \textit{Reverse Cowgirl anal}~\cite{Reverse-Cowgirl-anal}.

Finally, we examine the semantic relevancy of textual triggers to their corresponding models and target harmful services. Our analysis reveals that a large portion of triggers have limited semantic correspondence with either, implying that many triggers function as model activation tokens rather than explicit descriptions of model functionality or harmful purposes. We provide the detailed semantic analysis in Appendix~\ref{appendix:triggersemantics}.

\subsection{Harmful Services and Capabilities Provided by Monets}

\noindent\textbf{Harmful services advertised by Monets}.\label{subsubsec:categorization_results}
Using the harmful service classification framework in \S\ref{subsubsec:malicious_service_classification}, we identified 10 categories and 17 subcategories of harmful services advertised by Monets, according to their metadata. Among them, 14,921 Monets are labeled with more than one subcategory, and 41 belong to more than five subcategories.
As listed in Table~\ref{tab:taxonomy}, ``sex and nudity'' (A1-A5) is the most popular category---accounting for 75.74\% of all identified Monets---followed by ``intellectual property infringement'' (I1; 37.85\%), ``violence, gore, and horror'' (C1-C3; 3.91\%).

\noindent\textbf{Harmful capabilities provided by Monets}.\label{subsubsec:capabilities}
To validate whether identified Monets can generate harmful content as advertised, we sampled and evaluated them as follows.

\noindent$\bullet$\textit{ Model sampling}. 
We used stratified sampling to cover all harmful service subcategories. For each subcategory, we sampled 11 Monets from those that disclose their base models, obtaining 170 unique Monets in total\footnote{Subcategories D1 and H1 contain fewer than 11 eligible models (see Table~\ref{tab:taxonomy}), so all available eligible models in these subcategories were tested.}. We also included 18 corresponding base T2I models (see \S\ref{subsubsec:basemodel}) for comparison.

\noindent$\bullet$\textit{ Dataset}. We used the T2isafety dataset, a benchmark dataset that includes various categories of harmful T2I model prompts~\cite{li2025t2isafety}. Based on the harmful service subcategories of Monets (see Table~\ref{tab:taxonomy}), we extracted two prompts per subcategory for a total of 34 harmful prompts.
To evaluate the capabilities of Monets with developer-specified triggers, we tested each Monet under two settings: prompts with and without triggers. In the former, the trigger sequence specified by its developer was prepended to each prompt. Note that none of the validated base T2I models---which are general-purpose checkpoints---declare developer-specified triggers.

\noindent$\bullet$\textit{ Metrics}. 
To systematically assess Monets' harmful capability, we evaluated generated images in three dimensions: 
(1) \textit{alignment}, measuring whether outputs align with input prompts, scored by an alignment score~\cite{sahili2025fairjudge}; 
(2) \textit{quality}, assessing perceptual visual quality in terms of structure, color, sharpness, and noise, quantified by a quality score~\cite{wu2024comprehensive}; 
and (3) \textit{harmfulness}, judging the harmfulness of images, measured by a harmfulness score~\cite{ying2026safebench}. 
Following prior work~\cite{sahili2025fairjudge,wu2024comprehensive,ying2026safebench}, all scores range from 1 to 5 and are assigned by Qwen2.5-VL as MLLM-as-a-judge. We validate the MLLM-as-a-judge approach in Appendix~\ref{appendix:LLM-as-A-Judge}.

\noindent$\bullet$\textit{ Results}.
As shown in Table~\ref{tab:generated-image-score-comparison}, Monets achieve overall generation performance comparable to their corresponding base T2I models when evaluated across all sampled prompts. Specifically, ``Monets (All)'' perform similarly to the base models across all metrics, suggesting that Monets largely retain the general generation capability of their base models.

When evaluated on prompts associated with their target harmful services (i.e., ``Monets (Target)''), however, Monets outperform the base T2I models overall across all metrics. The improvement is particularly in harmfulness, where the average score increases from 3.07 for the base models to 3.42 without triggers and 3.49 with triggers. These results indicate that Monets are effectively specialized for generating content within their target harmful domains.

Developer-specified textual triggers further amplify this specialization. For ``Monets (Target),'' enabling triggers further increases image quality and harmfulness, indicating that these trigger sequences activate model behaviors associated with the harmful services.

Interestingly, triggers also influence generation across all sampled prompts. For ``Monets (All),'' enabling triggers improves image quality (3.83 to 4.00) but reduces prompt alignment (2.91 to 2.66), suggesting that triggers steer generation toward the models' learned styles or target domains even when prompts fall outside their target harmful services.

\begin{table}[t!]
\centering
\scriptsize
\caption{Average scores of images generated by Monets and their base models.}
\label{tab:generated-image-score-comparison}
\begin{threeparttable}
\begin{tabular}{lcccc}
\toprule
\textbf{Models} & \textbf{Trigger}\tnote{*} & \textbf{Alignment} & \textbf{Quality} & \textbf{Harmfulness} \\
\midrule
Base T2I models & \no & 2.82 & 3.73 & 3.07 \\
Monets (All) & \no & 2.91 & 3.83 & 3.12 \\
Monets (All) & \yes & 2.66 & 4.00 & 3.16 \\
Monets (Target)\tnote{**} & \no & 3.06 & 3.88 & 3.42 \\
Monets (Target) & \yes & 3.01 & 4.04 & 3.49 \\
\bottomrule
\end{tabular}
\begin{tablenotes}
\item[*] \yes\, and \no\, indicate prompts with and without triggers, respectively.
\item[**] Monets (Target) includes only outputs generated from prompts corresponding to each Monet's target harmful service category. 
\end{tablenotes}
\end{threeparttable}
\end{table}

\begin{findingbox}
\textit{\textbf{Takeaway 3.}} 
Monets are classified into 17 subcategories of harmful services, with ``sex and nudity'' as the dominant category. 
Empirical evaluation confirms that Monets are specialized for harmful image generation, outperforming their base models most notably in harmfulness within their target domains.
\end{findingbox}

\subsection{Technical Foundations of Monets}

\begin{figure}[t]
    \centering
    \includegraphics[width=0.8\linewidth]{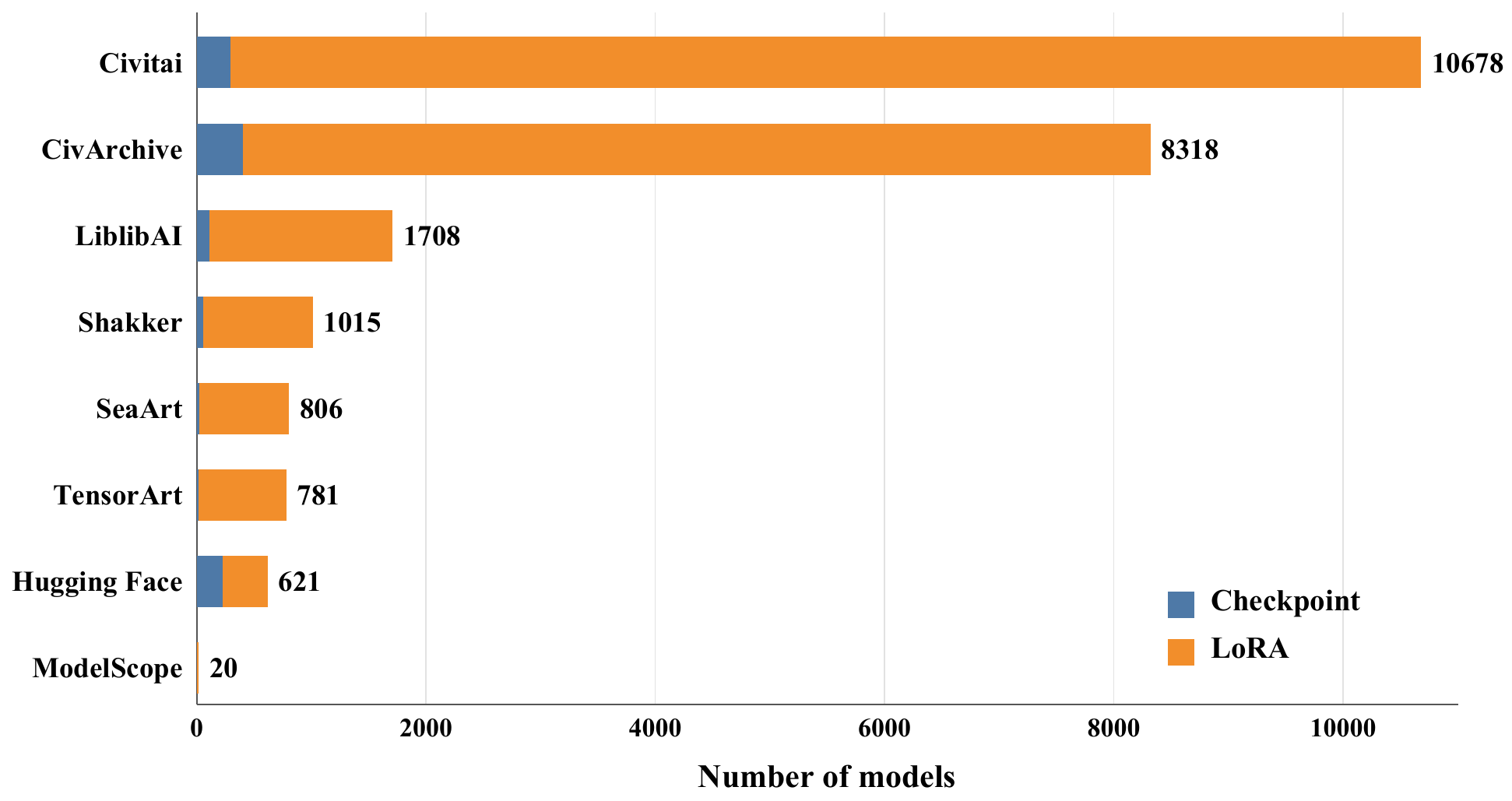}
    \caption{Model types across eight T2I model hubs.}
    \label{fig:model-type}
\end{figure}

\noindent\textbf{Model types and development methods}.
The collected Monets comprise 1,141 checkpoint models and 22,806 LoRA models. 
Although LoRA models dominate the ecosystem, their prevalence varies across model hubs. As shown in Fig.~\ref{fig:model-type}, TensorArt has the highest proportion of LoRA-based Monets (98.21\%), while Hugging Face leads in checkpoint-based Monets (37.04\%).

We further identify development methods based on model types and explicit disclosures in model metadata. Specifically, Civitai and Hugging Face provide structured development information for some models, while developers may additionally disclose development methods in model descriptions. We also treat LoRA models as evidence of fine-tuning, as LoRA is a parameter-efficient fine-tuning method~\cite{hu2022lora,dong2026aurora}. Based on these signals, we identify the development methods of 23,163 Monets. Among them, 22,952 Monets are developed through model fine-tuning, 211 through merging---including 63 that explicitly report merging LoRA models into checkpoints.

\noindent\textbf{Base models}.\label{subsubsec:basemodel}
By examining model metadata, we found that 22,135 Monets (92.43\%) disclose their base models.
As listed in Table~\ref{tab:base-model} of Appendix~\ref{appendix:baseModel}, we categorize the disclosed base models by model family and generative paradigm (see \S\ref{subsec:t2i_models}). Specifically, Illustrious (7,917) is the most prevalent base-model family, followed by Stable Diffusion 1.x/2.x (4,812) and Pony (4,779). These families follow community conventions~\cite{civitai-compare-base-models,civitai_base-model_constants} adopted by major model hubs such as Civitai and TensorArt. For example, SDXL-derived families such as Illustrious, Pony, and NoobAI are treated as distinct base models in practice and are therefore listed separately from SDXL.
These findings show that Monet development relies heavily on a small number of widely adopted open-source base-model families.

\noindent\textbf{Training datasets}.\label{subsubsec:training_data} 
Information about training datasets can be obtained from the structured ``datasets'' field in model metadata or from unstructured context in model descriptions.

\noindent$\bullet$\textit{ Disclosures in structured ``datasets'' field}. We identified 26 models, all hosted on Hugging Face, that explicitly disclose their training datasets in model metadata, covering 22 distinct datasets.
Specifically, seven datasets are primarily associated with sexually explicit content, four of which are labeled as ``Not-For-All-Audiences'' by Hugging Face. Examples include the datasets \textit{e621-rising-v3-curated}~\cite{e621-rising-v3-curated} and \textit{e621-rising-v3-finetuner}~\cite{e621-rising-v3-finetuner}, disclosed by the Monet \textit{e621-rising-v3}~\cite{e621-rising-v3}. Also, three datasets focus on deepfake generation, while another three target intellectual-property characters. 
In almost all cases, the intended harmful services of a Monet can be traced back to the domains of its training datasets.

11 datasets disclose their data sources in their descriptions: four are generated using existing T2I models, two are compiled from curated image collections, and five are collected from online sources. For example, the dataset \textit{e621-rising-v3-curated}~\cite{e621-rising-v3-curated} combines content gathered from e621.net, gelbooru.com, safebooru.donmai.us, and rule34.xxx.

\noindent$\bullet$\textit{ Disclosures in model description context}. 
Unlike Hugging Face, other model hubs do not provide structured training-dataset disclosures. However, Monet developers occasionally reveal data sources in model descriptions. For example, \textit{Aviana-v1.0}~\cite{Aviana}, a ``real-person likeness/deepfake'' Monet, is tailored to generate images of a specific social-media content creator and states that its training data were collected from the target creator's OnlyFans, X, Instagram, and Reddit accounts.
Beyond this example, we identified 15 additional deepfake Monets that explicitly target specific content creators, reference these individuals in their model names, and state in their descriptions that social-media content from the targeted individuals was used as training data.

\begin{findingbox}
\textit{\textbf{Takeaway 4.}}
Monet development is supported by readily accessible training data, ranging from harmful datasets openly hosted on model hubs---some explicitly flagged for sensitive content---to social-media images collected from identifiable individuals, raising concerns regarding privacy, consent, and identity misuse.
\end{findingbox}

\section{Ecosystem Dynamics of Monets}\label{sec:campaign}

To expand model collections, model hubs mirror and incorporate models originally hosted on major hubs such as Civitai and CivArchive, as well as smaller ones such as Tungsten~\cite{Tungsten} and Yodayo~\cite{Yodayo}. 
Leveraging this cross-platform model propagation and other methods, Monets also attempt to evade the platform-level moderation and model-level guardrails.
Meanwhile, model developers employ various promotion and monetization strategies to increase the reach and profit of related businesses.
Overall, such dynamics foster the proliferation of Monets across the ecosystem.

\begin{table}[t!]
\centering
\scriptsize
\caption{Counts of Monets with labeled upstream sources across model hubs.
}
\label{tab:upstream}
\begin{threeparttable}
\begin{tabular}{llrr}
\toprule
\multicolumn{1}{c}{\textbf{Model Hub}} & \multicolumn{1}{c}{\textbf{Upstream}} & \multicolumn{1}{c}{\textbf{\# Monets}} & \multicolumn{1}{c}{\textbf{Share in Hub}} \\
\midrule
CivArchive & Civitai       & 5,728 & 68.86\% \\
CivArchive & SeaArt        & 5,222 & 62.78\% \\
CivArchive & Hugging Face  & 2,099 & 25.24\% \\
CivArchive & TensorArt     & 1,470 & 17.67\% \\
CivArchive & TensorFiles   & 209   & 2.51\% \\
CivArchive & Tungsten      & 166   & 2.00\% \\
CivArchive & CivitasBay    & 157   & 1.89\% \\
CivArchive & PixAI         & 143   & 1.72\% \\
CivArchive & TensorHub     & 58    & 0.70\% \\
CivArchive & Moescape      & 22    & 0.26\% \\
CivArchive & Yodayo        & 22    & 0.26\% \\
CivArchive & Shakker       & 12    & 0.14\% \\
\midrule
SeaArt     & Civitai       & 749   & 92.93\% \\
SeaArt     & LiblibAI      & 1     & 0.12\% \\
\midrule
TensorArt  & Civitai       & 297   & 38.03\% \\
TensorArt  & PixAI         & 5     & 0.64\% \\
TensorArt  & Hugging Face  & 1     & 0.13\% \\
TensorArt  & LiblibAI      & 1     & 0.13\% \\
TensorArt  & SeaArt        & 1     & 0.13\% \\
\midrule
LiblibAI   & --\tnote{*}        & 387   & 22.66\% \\
\bottomrule
\end{tabular}
\begin{tablenotes}
\item[*] LiblibAI only indicates whether a model is mirrored, but does not specify upstream platform.
\end{tablenotes}
\end{threeparttable}
\end{table}

\subsection{Propagation of Monets across Model Hubs}\label{subsec:propagation}

\noindent\textbf{Platform-led propagation}.\label{subsubsec:Platform-propagation}
To expand their model collection, some model hubs mirror model artifacts---including model cards and, in some cases, model files---from other platforms.

As shown in Table~\ref{tab:upstream}, we identified a total of 9,760 Monets on CivArchive, SeaArt, TensorArt, and LiblibAI that explicitly disclose their upstream sources or mirroring statements within metadata. 
Specifically, LiblibAI only indicates that a model is mirrored, without specifying its source, while CivArchive, SeaArt, and TensorArt provide explicit source-platform information for all mirrored models. 
Notably, as a cross-platform model aggregator (see \S\ref{subsec:t2i_models}), all the 8,318 Monets identified from CivArchive are mirrored from other hubs.

To characterize the upstream sources of cross-platform propagation, we identified source models from 13 distinct model hubs based on the metadata of mirrored Monets hosted on CivArchive, SeaArt, and TensorArt. Aligned with the observation in \S\ref{subsubsec:distribution_growth}, Civitai serves as the dominant source hub, contributing the largest number of mirrored Monets across all three platforms. This finding reveals Civitai's central role as a source hub in the platform-led propagation of Monets.

When posting upstream sources, different model hubs adopt different strategies. Specifically, SeaArt and LiblibAI provide only a label ``Model Source'' or ``Mirroring,'' while TensorArt exposes links to the original model cards hosted on the source platforms. As a large-scale model aggregator spanning multiple platforms, CivArchive provides the most comprehensive and sophisticated source information. Different from the above three hubs that typically post a single upstream source, CivArchive may associate a model with multiple upstream sources---with an average of 1.84 sources---and record them into two source fields (i.e., ``Available-on'' and ``Mirrors''). 
The ``Available-on'' field records links to model cards on upstream platforms, while the ``Mirrors'' field records direct download links to model files hosted on upstream platforms. 
As shown in Appendix Table~\ref{tab:civarchive-upstream-source-type}, all CivArchive Monets contain at least one ``Available-on'' entry, while 5,764 (69.30\%) include at least one ``Mirrors'' entry.
Notably, CivArchive adopts different source-disclosure strategies for distinct upstream platforms---for example, references to Hugging Face-hosted files appear exclusively in the ``Mirrors'' field---hinting at its cross-platform archival mechanisms further discussed in \S\ref{subsubsec:archiving}.

\noindent\textbf{Developer-led propagation}.\label{subsubsec:developer-led-propagation}
To promote Monets across multiple model hubs and maintain alternative access paths, developers often include \textit{model referral links} in model descriptions that direct users to the same model hosted on other hubs.

As shown in Appendix Table~\ref{tab:model-referral-link-in-description}, we collected 2,520 model referral links from 2,093 Monets carrying descriptions in metadata. Among model hubs, Civitai is the most frequently referenced hub, accounting for 47.86\% of all model referral links. 
TensorArt (19.40\%) and Hugging Face (12.34\%) are the next most frequently referenced hubs.

We also observe model referral links pointing to smaller T2I model hubs, such as Yodayo and TensorHub~\cite{tensorhub}, indicating that developers use a diverse set of platforms to increase model exposure and maintain cross-platform availability.

\begin{findingbox}
\textit{\textbf{Takeaway 5.}} 
Monets propagate cross-platform via two complementary mechanisms: platform-led mirroring (40.76\% of Monets) and developer-embedded model referral links (8.74\%), with Civitai acting as the central hub in both.

\end{findingbox}

\subsection{Governance Evasion}
In our study, we identified three governance evasion strategies employed by Monets, with two targeting platform moderation and one circumventing model guardrails.

\noindent\textbf{Cross-platform model archiving}.\label{subsubsec:archiving} 
Originally established to preserve access to models removed from Civitai or other hubs, both CivArchive and SeaArt explicitly archive and redistribute model files from multiple hubs at scale, including models that may violate platform policies~\cite{CivArchivepolicy,seaart-archive-civitai}. Such archival practices are especially common for Monets.

However, the two platforms differ in both infrastructure and scope: SeaArt directly mirrors model files on its own platform, while CivArchive operates a more sophisticated archival infrastructure involving external storage and cross-platform references. Moreover, unlike SeaArt, CivArchive publishes no content policy or takedown procedure on its website~\cite{CivArchivepolicy}, and its contribution guide imposes no restrictions on what may be indexed~\cite{CivitAIArchive_Uploading}---offering a more tolerant environment for harmful content.
These factors make CivArchive the most critical platform for archiving cross-platform Monets, as reflected in Table~\ref{tab:upstream}. 
We thus focus on CivArchive and examine its archival practices from three aspects: the archival mechanism, archival repositories, and banned-but-alive models.

\noindent$\bullet$\textit{ Archival mechanism}. Based on CivArchive's upload guidance~\cite{CivitAIArchive_Uploading}, Hugging Face serves as its primary archival platform for storing model files. CivArchive surfaces these archived files to users via ``Mirrors'' entries (see \S\ref{subsubsec:Platform-propagation}).
Further examining Hugging Face model files referenced by ``Mirrors'' entries across 2,099 CivArchive Monets that carry such references (see Table~\ref{tab:upstream}), we collected 3,820 unique mirrored model files hosted across 942 Hugging Face repositories, uploaded by only 424 unique accounts. This indicates that archiving is concentrated among a small group of repositories and users.

\noindent$\bullet$\textit{ Archival repositories}. Deeper analysis reveals a concentrated archival pattern in repositories. Among the 942 repositories, 74.84\% of them host model files that belong to multiple distinct models, and 34.39\% of the repositories each host over 100 model files. 
For instance, by matching file names, we found that the repositories \textit{loras}~\cite{RectalWorm-loras} and \textit{Zacygiz\_lora}~\cite{Zacygiz-lora} host an identical set of 9,982 files, approaching Hugging Face's 10k-per-folder ceiling~\cite{HF-storage-limit}, of which 365 are identified Monets. Such volumes indicate that archival uploads are performed at scale, consistent with CivArchive's upload guidance~\cite{CivitAIArchive_Uploading}.

Also, 80.47\% of these repositories either contain no model card (62.00\%) or provide only an empty model card (18.47\%). Given that Hugging Face relies heavily on user reporting for moderation decisions~\cite{HFpolicy}, the absence of meaningful model descriptions may reduce the visibility of harmful models to both users and moderators. Consistent with this observation, only 11.68\% of repositories containing Monet model files are labeled as ``Not For All Audiences''~\cite{HFpolicy} by Hugging Face.

\noindent$\bullet$\textit{ Banned-but-alive models}.
Among the Monets archived on CivArchive, 997 (11.99\%) of them, all originating from Civitai, had already been removed from their original platform yet remain publicly accessible through CivArchive. This reveals the limitation of platform-siloed moderation, as Monets can persist ecosystem-wide via cross-platform archiving even after being banned on their original platforms.

\begin{findingbox}
\textit{\textbf{Takeaway 6.}}
CivArchive preserves Monets via a centralized archival infrastructure concentrated in a small number of Hugging Face repositories, most of which lack meaningful model cards, potentially reducing their visibility to moderation mechanisms. 
Consequently, 11.99\% of Monets archived on CivArchive had already been banned from their original platforms yet remain publicly accessible, highlighting the limitations of platform-siloed moderation. 
\end{findingbox}

\noindent\textbf{Keyword obfuscation}. To circumvent keyword-based moderation mechanisms, some Monets employ keyword obfuscation~\cite{yang2021scalable} in model metadata. To explore this behavior, we manually reviewed the names of the collected Monets and identified 66 models containing obfuscated representations of sensitive terms.
The techniques include symbol insertion, character substitution, and word separation. Examples include ``F*CK'' (i.e., ``FUCK''), ``pub1ch41r'' (i.e., ``pubic hair''), and ``P\_or\_n'' (i.e., ``Porn'')~\cite{middle-finger-fck,PubicHair,Pornconceptz}, which are related to sexually explicit content. We also observed obfuscated violence-related terms such as ``M*rder'' (i.e., ``Murder''), ``CR1M3'' (i.e., ``Crime''), and ``Su1cide'' (i.e., ``Suicide'')~\cite{murder_investigator,CR1M3,suicide_squad}.

\begin{findingbox}
\textit{\textbf{Takeaway 7.}} 
Some Monet developers actively obfuscate sensitive keywords in model metadata to evade keyword-based moderation, revealing that such moderation mechanisms are insufficient against adversarial developers.
\end{findingbox}

\noindent\textbf{Limited adoption and circumvention of model-level safeguards}.\label{subsubsec:safetyCheckerCircumvention} 
Model-level safeguards are safety modules distributed along with model files, of which the Safety Checker~\cite{DiffusionPipeline,safetychecker} is the most widely accessible: an optional yet official safety-filtering module integrated into Hugging Face's \textit{diffusers} library, designed to detect and filter harmful content generated by diffusion-based T2I models. 
However, only 45 checkpoint-based Monets include this module, eight of which explicitly provide instructions for disabling it via example scripts in their model descriptions (e.g., \textit{sexy\_toon\_3d\_moresexy}~\cite{sexytoon3dmoresexy}).

\begin{findingbox}
\textit{\textbf{Takeaway 8.}} 
Model-level safety mechanisms are sparsely adopted among Monets, with some developers explicitly providing instructions to disable them, highlighting the limitations of relying on developer-controlled safeguards for ecosystem-wide protection.
\end{findingbox}

\subsection{Promotion and Monetization of Monets}\label{subsec:monetization}

By analyzing Monet metadata, we found that developers frequently embed \textit{external links} (i.e., links referring to external platforms and contact information), facilitating cross-platform promotion and monetization of Monet-related products and services.
To understand these promotion and monetization practices, we systematically analyzed the destinations and purposes of external links stated in Monet metadata.

\noindent\textbf{Categorization of external links}. From Monet metadata, especially descriptions, we extracted cross-platform links. After removing the model referral links that were discussed in \S\ref{subsubsec:developer-led-propagation}, we extracted a total of 1,393 external links from 5,405 Monets and categorized them into seven types based on platform characteristics: digital payment platforms, paid-access platforms, social media platforms, art-sharing platforms, messaging channels, link aggregators, as well as external documents and websites. 
Among them, paid-access platforms dominate (43.29\%), followed by social media and messaging channels.

\noindent\textbf{Cross-platform promotion and monetization strategies}. Different types of external links support distinct promotion and monetization strategies, as detailed below.

\noindent$\bullet$\textit{ Digital payment platforms}. As the most direct form of monetization, some developers embed payment links (e.g., PayPal and Alipay) in model descriptions to receive donations.

\noindent$\bullet$\textit{ Paid-access, social-media, and art-sharing platforms}. Aiming to showcase model capabilities and attract potential users, developers often promote AI-generated images across paid-access platforms (e.g., Buy Me a Coffee~\cite{buymeacoffee}, Ko-fi~\cite{Kofi}, and Patreon~\cite{patreon}), social-media platforms (e.g., Facebook and X), and art-sharing platforms (e.g., Pixiv~\cite{pixiv} and DeviantArt~\cite{deviantart}). 

Unlike promotion-oriented platforms, paid-access platforms enable direct monetization via purchases, subscriptions, and memberships.
On paid-access platforms, beyond images, developers also monetize related products and services, like prompts, premium models, and model-training services. For instance, the developer of \textit{Murky's PDXL Lite}~\cite{MurkysPDXLLite-civitai} publishes a lightweight version on Civitai while offering the full version on Patreon~\cite{MurkySkeleton}. Similarly, the developer of \textit{Arabatos LoRA-v1.0}~\cite{ArabatosLoRA} advertises a LoRA training service and directs users to Ko-fi~\cite{ownwaifu}, where LoRA models and images are sold, with over 3K payments recorded.

\noindent$\bullet$\textit{ Link aggregators}. Link aggregators (e.g., Linktree~\cite{linktree}) play an infrastructural role by consolidating links to paid-access platforms, social-media accounts, art-sharing platforms, or more, facilitating user redirection between multiple platforms.

\noindent$\bullet$\textit{ Messaging channels, external documents, and websites}. 
These types of links reveal more complex ecosystem interactions. 
Specifically, we identified collaborative documents linked from 1,109 Monet descriptions, through which users can request new datasets or models.
Also, 163 models link to cloud-hosted model operation infrastructures, while 86 link to model-usage instructions. Notably, 40 models contain links or contacts associated with gray-market services, all related to AI-assisted account creation and farming.

\noindent\textbf{Campaign analysis}. \label{subsubsec:campaign-analysis}
While examining these external links embedded in Monet metadata, we observed that developers frequently reuse the same external links across multiple Monets. In some cases, these external links are displayed on different developer accounts and platforms, revealing coordinated promotion activities.
This behavior resembles patterns observed in other online campaigns, where shared infrastructure and contact information are reused across multiple entities or activities~\cite{roundy2020many,wang2022demystifying}. Motivated by this observation, we adopted the principle of guilt by association (GBA)~\cite{wang2017gang} to explore the campaigns operating behind Monets.

We built a heterogeneous graph with three types of nodes: Monets, external links (see \S\ref{subsec:monetization}), and developer accounts. Each Monet is connected to its developer account and the external links disclosed in its metadata. For model referral links (see \S\ref{subsubsec:developer-led-propagation}), we directly connect the Monet containing the referral link to the Monet referenced by that link. In this way, Monets are linked within the graph through shared developer accounts, shared external links, or direct model referrals.

The resulting graph contains 6,879 connected components. After excluding components in which Monets are connected solely via shared developer accounts, 398 components contain at least two Monets, collectively covering 5,090 models. Following the GBA principle, we treat each remaining connected component as a campaign. The validation of this campaign identification approach is detailed in Appendix~\ref{appendix:campaign-validation}.

\begin{figure}[t!]
    \centering
    \includegraphics[width=0.8\linewidth]{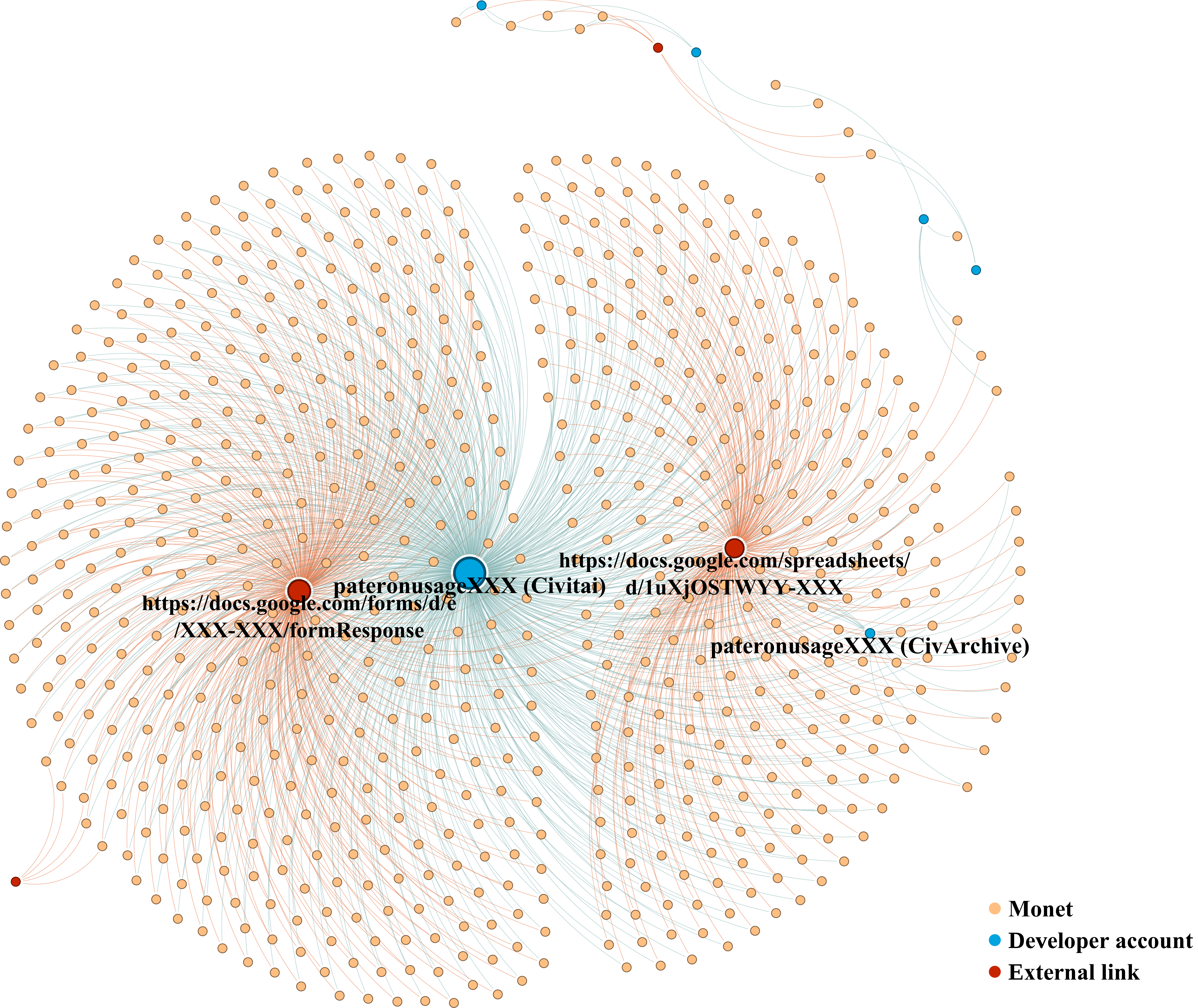}
    \caption{The campaign subgraph of Case 1.}
    \label{fig:case1_campaign}
\end{figure}

\noindent$\bullet$\textit{ Case 1: A campaign mainly connected through external documents}. The largest connected component contains 668 Monets spanning three model hubs: CivArchive, Civitai, and Shakker. These Monets primarily focus on pornography and intellectual-property characters.
Significantly, 98.50\% of these Monets (e.g., \textit{Ikumi Mito (Food Wars) - LoRA Illustrious [NSFW Support]}~\cite{Ikumi_Mito}) are connected through two shared external documents, as shown in Fig.~\ref{fig:case1_campaign}. The first is a Google Forms survey~\cite{Ikumi_Mito-survey} listing 649 candidate characters and art styles for future LoRA development, together with corresponding commission prices, through which users can request new models. The second is a Google Doc~\cite{Ikumi_Mito-commissions} showcasing 914 completed commissions. The shared resources indicate an organized commission-based model-development campaign operating across multiple hubs.

\begin{figure}[t!]
    \centering
    \includegraphics[width=0.78\linewidth]{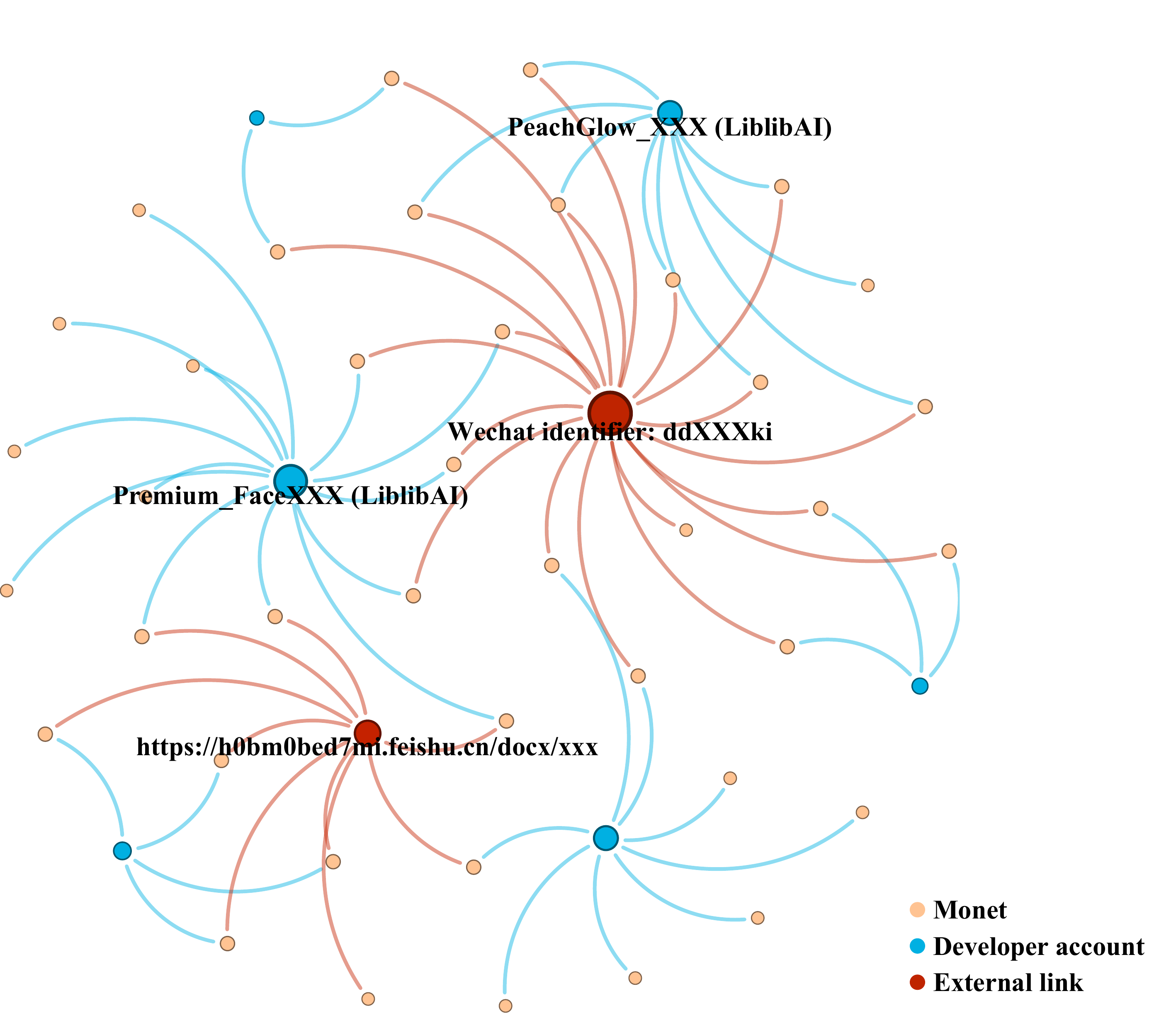}
    \caption{The campaign subgraph of Case 2.}
    \label{fig:case2_campaign}
\end{figure}

\noindent$\bullet$\textit{ Case 2: A campaign mainly connected through gray-market services}.\label{subsubsubsec:account_farming}
We also identified a connected component consisting of 40 Monets, from LiblibAI and Shakker, whose metadata advertise an AI-assisted account-farming service. According to the service description, it aims to facilitate the large-scale creation of social-media accounts, attract followers, and monetize the resulting traffic via advertising activities. The target social media include TikTok and RedNote~\cite{Xiaohongshu}, both of which explicitly prohibit such account-farming activities under their platform policies~\cite{tiktok-ToS,Xiaohongshu-ToS}.
As shown in Fig.~\ref{fig:case2_campaign}, the models in this component are linked via shared contacts, including identical WeChat identifiers, links to the same service website~\cite{Feishu_case}, as well as developer accounts. Notably, the Monets used to promote this service are associated with pornography and deepfake generation. The reuse of shared advertisements and contacts suggests a coordinated effort to leverage Monets as a channel for advertising and acquiring customers for related commercial services.

\begin{findingbox}
\textit{\textbf{Takeaway 9.}} 
Beyond individual model distribution, Monets are organized by coordinated cross-platform campaigns: one commission-based campaign spans 668 Monets across three hubs with 914 completed orders, while another uses pornographic or deepfake Monets to advertise gray-market account-farming targeting TikTok and RedNote, revealing a sophisticated commercial ecosystem behind Monets.
\end{findingbox}

\section{Downstream Monet Deployment}\label{sec:deploy}

\begin{table}[t!]
\centering
\scriptsize
\setlength{\tabcolsep}{1.9pt}
\caption{Categories of Monet-integrated GitHub projects.}
\label{tab:projects}
\begin{tabular}{clr|clr}
\toprule
\textbf{Rank} & \multicolumn{1}{c}{\textbf{Category}} & \textbf{Count} & \textbf{Rank} & \multicolumn{1}{c}{\textbf{Category}} & \textbf{Count} \\

\midrule
1 & Node workflow systems         & 267 &
9 & Video/animation creation  & 28 \\

2 & Notebooks/hosted demos      & 101 &
10 & Model management              & 26 \\

3 & Model training/fine-tuning  & 90  &
11 & Web/API services & 22 \\

4 & Image generation/editing    & 79 &
12 & Data scraping/analysis      & 13 \\

5 & Infrastructure                & 65  &
13 & Domain-specific app.  & 10 \\

6 & Documentation/websites & 59 &
14 & Model repositories           & 6 \\

7 & Research/benchmark          & 48  &
15 & Security governance tools     & 6 \\

8 & Chatbot/agent/RAG         & 42  &  \\

\bottomrule
\end{tabular}
\end{table}

To examine downstream deployment of Monets, we investigated their use in GitHub projects and AI API providers. 

\subsection{Monet-Integrated GitHub Projects}

\noindent\textbf{Data collection}.
To discover GitHub projects that integrate Monets, we queried GitHub's code search engine using the model filenames and identifiers of the collected Monets. 
In this way, we identified 1,102 repositories that integrate 216 Monets. The distributions of these Monets by their source model hubs and target harmful services are listed in Appendix Table~\ref{tab:metadata-hub-distribution} and Table~\ref{tab:model-category-distribution}, respectively.

\noindent\textbf{Project functionality}.
To understand project functionality, two experts manually annotated the collected Monet-related projects by reviewing README files and Monet-included scripts, classifying them into 15 categories (see Table~\ref{tab:projects}). Projects were categorized only when both annotators agreed, yielding Cohen's $\kappa=0.88$.

Due to incomplete or absent project descriptions, we are unable to categorize 240 projects. Among the categorized projects, the category of node workflow systems~\cite{xue2025comfybench} dominates with 267 projects---generative AI systems using visual node-based graphs as the primary interface for constructing and executing AI workflows (e.g., ComfyUI)---followed by 101 projects for Notebooks and demonstrations and 90 for model training and fine-tuning. 
Note that only six repositories host Monet model files, consistent with prior findings~\cite{ait2025suitability} that GitHub does not serve as a major platform for hosting AI models within the model-sharing ecosystem.

Among all collected projects, we identified 15 designed for harmful services, including eight for pornographic and intellectual-property character image generation, five for NSFW chat, and two for account-farming services, collectively adopting 20 Monets identified in this study. Pornographic and intellectual-property character image generation projects act as single-task tools, while NSFW chat projects often construct more complex platforms that integrate both text generation and image creation,
deployable locally and via popular messaging applications such as Telegram (e.g., \textit{ClawFans}~\cite{ClawFans}) and WhatsApp (e.g., \textit{BBA}~\cite{bba}). The account-farming projects, related to the gray-market services discussed in \S\ref{subsubsubsec:account_farming}, automate the creation, management, and deployment of AI influencers, including persona construction, content generation, and social media presence (e.g., \textit{Gator}\cite{gator}).

\noindent\textbf{Case Study: ClawFans}.
ClawFans~\cite{ClawFans} is an AI-powered character chat platform project that supports NSFW interactions and can be integrated with Telegram as a chatbot. It employs \textit{Qwen2.5-14B-Instruct-abliterated~}\cite{Qwen2.5-14B-Instruct-abliterated} for text generation and \textit{NoobAI-XL-Vpred-v1.0}~\cite{NoobAI-XL-CivArchive,NoobAI-XL-Civitai} for image generation.
Notably, the former is an uncensored LLM known to generate harmful content without conventional safety guardrails~\cite{lin2025consiglieres}, while the latter is identified as a Monet in our study and also tagged as ``NSFW'' in its metadata~\cite{NoobAI-XL-CivArchive}, intended for NSFW anime image creation as stated in both its metadata and the ClawFans repository~\cite{ClawFans-download_noobai}.

As an NSFW-oriented role-playing platform, ClawFans includes numerous virtual character profiles. Among these, we identified character profiles that portray minor individuals, including one explicitly described as a 17-year-old girl~\cite{ClawFans-seed_characters}. The combination of an uncensored LLM and a NSFW Monet in such a NSFW role-play setting raises concerns about the potential generation of sexualized content involving minors and associated risks of child sexual abuse material.

\begin{findingbox}
\textit{\textbf{Takeaway 10.}} 
Monets are integrated into 1,102 GitHub projects, with 15 explicitly designed for harmful services, including pornographic and intellectual-property character image generation, NSFW chat, and account-farming. Critically, some downstream projects combine uncensored LLMs with NSFW Monets in role-play settings involving minor characters, raising serious child safety concerns.
\end{findingbox}

\subsection{Monet APIs from AI API Providers}
\noindent\textbf{Data collection}. 
To characterize Monets adopted by AI API providers, we examined two popular AI API providers that explicitly expose T2I inference services: Runware~\cite{runware} and SogniAI~\cite{sogni}.
We queried each provider's model search engine and model catalog using model names and identifiers, identifying a total of 3,930 Monet API entries. Runware hosts 3,928 matched Monet APIs, while SogniAI contributes two.

\noindent\textbf{Adoption of Monets by AI API providers}.
We analyzed adopted Monets from three dimensions: source model hubs, tailored harmful services, and popularity.

As shown in Appendix Table~\ref{tab:metadata-hub-distribution}, the majority of the adopted Monets originate from Civitai (2,302; 58.58\%) and CivArchive (1,517; 38.60\%), indicating that API providers rely heavily on the dominant hubs of the Monet ecosystem (see \S\ref{subsubsec:distribution_growth}).
Since a Monet may support multiple harmful services (see \S\ref{subsubsec:malicious_service_classification}), the 3,930 adopted Monets collectively provide 7,264 harmful services spanning 16 subcategories (see Appendix Table~\ref{tab:model-category-distribution}). ``Sexualized posing, angles, and props'' (A4) is the most prevalent (2,738 Monets), followed by ``intellectual-property infringement'' (I1; 1,775) and ``explicit organs, fluids, and fetishes'' (A3; 1,263).

We further examined their popularity on original model hubs. Among the 3,930 adopted Monets, 202 (5.14\%) have accumulated over 10,000 downloads. For example, \textit{NSFW POV All}~\cite{NSFW-POV-All}, associated with sexual and pornographic content, has exceeded 203K downloads. Note that since API providers do not disclose usage volumes, we cannot measure the actual usage of these Monet APIs.

\begin{findingbox}
\textit{\textbf{Takeaway 11.}} 
Monets have propagated beyond model-sharing platforms into commercial AI inference services. By exposing Monets through hosted APIs, these providers allow users to access harmful image-generation capabilities without downloading or deploying models locally.
\end{findingbox}

\section{Discussion}\label{sec:discussion}

\noindent\textbf{Mitigation}.\label{subsec:mitigation}
As described above, our findings demonstrate that Monets are supported by an interconnected ecosystem spanning model development, cross-platform distribution, governance evasion, as well as monetization and downstream deployment.
Therefore, beyond interventions to individual models or platforms, we discuss ecosystem-wide mitigation against Monet abuse from four dimensions.

\noindent$\bullet$\textit{ Securing resources for model development}.
Our analysis in \S\ref{subsubsec:training_data} shows that Monet development can leverage readily accessible harmful training data, including sensitive datasets hosted on model hubs and images of identifiable individuals collected from social media. Therefore, safeguards should move upstream from model moderation to the resources used for model development. Platforms hosting datasets (e.g., Hugging Face) should strengthen dataset moderation through content screening, provenance documentation, and access controls for sensitive datasets. 
Similarly, cloud-based model-training platforms should vet both training data and resulting models before public release.

\noindent$\bullet$\textit{ Cross-platform moderation for model propagation}.
Our measurements show that platform-siloed moderation is insufficient when models are routinely mirrored, referred, and archived across platforms. For example, the large-scale migration to CivArchive following Civitai's policy update (see \S\ref{subsubsec:distribution_growth}) and the continued availability of Monets removed from their original hubs (see \S\ref{subsubsec:archiving}) demonstrate how harmful models can persist despite platform-specific enforcement. Therefore, model hubs should coordinate moderation across platforms by sharing identifiers or hashes of policy-violating models, tracking their upstream and mirrored copies, and distributing policy-removal signals across platforms.

\noindent$\bullet$\textit{ Enforcing safety at model and inference levels}.
Our findings in \S\ref{subsubsec:safetyCheckerCircumvention} reveal that model-level safeguards cannot be assumed to remain enabled: only a small number of Monets carry Safety Checkers, and some explicitly provide instructions for disabling them. Therefore, rather than relying solely on safeguards voluntarily retained by model developers, safety protection could be independently enforced by model hubs and API providers within their hosted inference pipelines.
We further examine existing technical building blocks for such inference-time enforcement, including Safety Checker~\cite{safetychecker}, Ethical-Lens~\cite{cai2025ethical}, SteerDiff~\cite{zhang2024steerdiff}, SAFREE~\cite{yoon2025safree}, and STG~\cite{na2026training}, which apply safety interventions at different stages of the generation pipeline (see Appendix~\ref{appendix:mitigationTools}). These mechanisms provide practical building blocks for platform-enforced safeguards that operate independently of model developers and are harder to disable than safety modules distributed together with model files.

\noindent$\bullet$\textit{ Extending governance to monetization and downstream deployment}.
The Monet ecosystem extends beyond model hubs. As shown in \S\ref{subsec:monetization}, developers leverage external platforms and coordinated campaigns to sell models and their created images, offer model-training services, and advertise gray-market services. \S\ref{sec:deploy} shows that Monets are integrated into GitHub projects and exposed via commercial AI APIs, enabling access to harmful capabilities without local model deployment. Therefore, the governance of Monets should extend to downstream services and monetization channels. Downstream projects and API providers should screen models before onboarding, retain upstream provenance, and enforce inference-time safeguards. Model hubs should also investigate recurring referral links, payment channels, and other shared infrastructure associated with coordinated harmful campaigns.

\noindent\textbf{Limitations}. Our study has two main limitations.
First, our discovery pipeline has two inherent constraints. Due to the large number of models across numerous model hubs, we relied on keyword-based retrieval on eight major hubs to identify candidate Monets, which may miss stealthy models that avoid harmful keywords or are hosted on non-primary hubs. Additionally, our metadata-based classification may not detect models that intentionally conceal harmful intent. Thus, the Monets identified in our study should be regarded as a lower bound of the true Monet population in the open-source T2I ecosystem. Nevertheless, these models represent the most explicitly advertised and publicly accessible Monets, making them the most user-discoverable and governance-actionable portion of the ecosystem.
Second, given the scale of identified Monets and the substantial cost of large-scale image creation and evaluation, we assessed harmful capabilities in \S\ref{subsubsec:capabilities} via stratified sampling instead of exhaustive testing.
This ensures coverage of all harmful service subcategories. 
Since this evaluation aims to validate whether Monets generate harmful content as advertised, stratified sampling across subcategories provides sufficient coverage without exhaustively testing individual models.

\section{Conclusion}
In this work, we present the first systematic measurement study of the Monet ecosystem. We uncover the prevalence and usage of Monets, shedding light on as many as 23,947 Monets from eight major T2I model hubs. In particular, we examine the real-world Monet ecosystem from five dimensions---model characteristics, cross-platform propagation, governance evasion, monetization, and downstream deployment. 
Additionally, our study reveals that Monets pose significant security threats embedded within this ecosystem: they persist across platforms despite moderation actions, operate coordinated gray-market campaigns, and are deployed in downstream projects raising potential child safety concerns.
Our findings provide new insights into the Monet ecosystem and inspire future efforts toward more robust governance and safety mechanisms to secure the open-source T2I ecosystem as a whole.

\bibliographystyle{plainurl}
\bibliography{refs}

\appendix
\section*{Appendix}

\section{False Case Analysis}

For T2I model checking, we observed no false positives among the sampled models. The false negatives primarily arise when models implicitly indicate their T2I capabilities without explicitly stating that they support image generation. For example, the Monet \textit{Pantie around ankles V2}~\cite{Pantiearoundankles} describes itself as a LoRA designed for Z-Image-family checkpoints, while \textit{Graphics Design - Product Ad}~\cite{GraphicsDesign} is described as a LoRA for FLUX.1 checkpoints. Although both belong to T2I model families, neither explicitly states its T2I generation capability, causing them to be missed by our checker.

For harmful service classification, false positives often occurred when the LLM misinterpreted the meaning or context of statements in model descriptions, especially explicit warnings, prohibitions, and copyright disclaimers. For example, the description of \textit{Real Dream}~\cite{Real-Dream} explicitly discourages its use for political manipulation, but the LLM-based classifier misinterpreted this warning as support for such use and classified the model under the ``Political misinformation'' subcategory. In another example, a copyright disclaimer in the description of \textit{Hinatazaka46}~\cite{Hinatazaka46} was mistakenly treated as evidence of ``Intellectual property infringement.''

False negatives mainly resulted from limitations in entity recognition, particularly for ``Real-person likeness/deepfakes'' and ``Intellectual property infringement'' subcategories. The LLM could miss references to real persons, groups, or copyrighted entities, especially when aliases, uncommon names, or character-level obfuscation were used. For example, the model \textit{HAER1N NEWJE4N5}~\cite{HAER1N-NEWJE4N5} obfuscates ``HAERIN,'' the name of a member of the South Korean girl group NewJeans, and ``NEWJEANS'' through digit substitution. Although readily recognizable to human reviewers, these references were missed by the LLM-based classifier, resulting in a false negative for intellectual-property infringement.

\section{Details of Semantic Analysis on Triggers}\label{appendix:triggersemantics}

To investigate whether textual triggers semantically reflect the corresponding models and their target harmful services, we measure the semantic similarity between each trigger and two types of contextual information: (1) the corresponding model name and description and (2) the description of the harmful service targeted by the model. Specifically, we encode the triggers and contextual information using \textit{paraphrase-multilingual-MiniLM-L12-v2}~\cite{sentencebert-model} and compute their pairwise cosine similarity. The results are shown in Table~\ref{tab:trigger_similarity}. We find that 55,228 textual triggers (27.26\% of the analyzed triggers) exhibit low semantic similarity to both the corresponding model information and target harmful services, with similarity scores below 0.2. 
For example, one Shakker developer~\cite{killerModel-shakker} who primarily publishes pornographic Monets uses the same developer identifier---unrelated to pornography---as a textual trigger across 74 of their Monets. 
These findings suggest that in the real world, a substantial portion of textual triggers serve primarily as model-specific activation tokens rather than semantically meaningful descriptions of the corresponding models or their target harmful services.

\begin{table}[t!]
\centering
\scriptsize
\color{black}
\caption{Distribution of semantic similarity between triggers and the corresponding models.}
\label{tab:trigger_similarity}
\renewcommand{\arraystretch}{1.0}
\begin{threeparttable}
\begin{tabular}{crrr}
\toprule
\textbf{Similarity Range} &
\textbf{\makecell{Model name\\\& description}} &
\textbf{\makecell{Target service\\description}}  \\
\midrule
$0.9 \leq x \leq 1.0$   & 0      & 0       \\
$0.8 \leq x < 0.9$   & 53     & 0       \\
$0.7 \leq x < 0.8$   & 496    & 1       \\
$0.6 \leq x < 0.7$   & 1,976  & 105    \\
$0.5 \leq x < 0.6$   & 5,260  & 2,376   \\
$0.4 \leq x < 0.5$   & 13,515 & 11,735  \\
$0.3 \leq x < 0.4$   & 34,513 & 25,133  \\
$0.2 \leq x < 0.3$   & 59,148 & 53,605  \\
$0.1 \leq x < 0.2$   & 59,006 & 57,899  \\
$0.0 \leq x < 0.1$   & 25,584 & 35,942  \\
$-0.1 \leq x < 0.0$  & 3,002  & 14,824  \\
$-0.2 \leq x < -0.1$ & 62     & 980     \\
$-0.3 \leq x < -0.2$ & 2      & 17      \\
$x < -0.3$ & 0      & 0      \\
\bottomrule
\end{tabular}
\begin{tablenotes}
\item[*] For a Monet assigned to multiple harmful subcategories, the service similarity score is calculated as the maximum cosine similarity between the trigger and the definitions of all assigned harmful subcategories.
\end{tablenotes}
\end{threeparttable}
\end{table}

\section{Details of Monet Base Models}\label{appendix:baseModel}

Table~\ref{tab:base-model} summarizes the distribution of base-model generative paradigms and families underlying the collected Monets. Denoising diffusion models constitute the majority. Flow-matching models account for a smaller but notable portion, dominated by the FLUX.1 family. For a small portion of Monets, the model metadata does not disclose their base-model generative paradigms or families. Thus, we categorize them as ``Unknown.''

\begin{table}[t!]
\scriptsize
\centering
\caption{Distribution of base-model generative paradigms and families.}
\label{tab:base-model}
\begin{tabular}{llr}
\toprule
\textbf{Generative Paradigm} & \textbf{Family} & \textbf{\# Monets} \\
\midrule
\multirow{9}{*}{Denoising Diffusion}
& Illustrious       & 7,917 \\
& Stable Diffusion 1.x/2.x  & 4,812 \\
& Pony              & 4,779 \\
& SDXL              & 1,341 \\
& NoobAI            & 209 \\
& Cosmos-Predict2   & 80 \\
& PixArt            & 3 \\
& HunyuanDiT        & 1 \\
\midrule
\multirow{7}{*}{Flow Matching}
& FLUX.1            & 1,844 \\
& Z-Image           & 830 \\
& Qwen-Image      & 148 \\
& FLUX.2    & 116 \\
& Chroma          & 39 \\
& Stable Diffusion 3.5           & 11 \\
& HiDream         & 4 \\
& HunyuanImage    & 1 \\
\midrule
Unknown & Unknown & 1,812 \\
\bottomrule
\end{tabular}
\end{table}

\section{Validation of MLLM-as-a-Judge Approach Used in Harmful Capability Evaluation}
\label{appendix:LLM-as-A-Judge}

As mentioned in \S\ref{subsubsec:capabilities}, we adopted a MLLM-as-a-judge approach powered by Qwen2.5-VL to evaluate the alignment, quality, and harmfulness of images generated by Monets. To further validate this approach, we evaluated the agreement between the MLLM-as-a-judge and human evaluators.

Specifically, we invited two security experts to independently annotate 100 randomly sampled images generated by the T2I models evaluated in \S\ref{subsubsec:capabilities} (excluding images of Subcategory A1). The annotators achieved Cohen's $\kappa$ values of 0.74, 0.79, and 0.76 for alignment, quality, and harmfulness, respectively, indicating strong agreement. Disagreements were resolved through discussion to establish the ground truth.

Compared with this ground truth, Qwen2.5-VL achieved Cohen's $\kappa$ values of 0.72, 0.74, and 0.75 for alignment, quality, and harmfulness, respectively, indicating strong agreement with human judgment.

\section{Supplementary Details of Monet Cross-Platform Propagation}

Tables~\ref{tab:civarchive-upstream-source-type} and~\ref{tab:model-referral-link-in-description} provide supplementary statistics for the platform-led and developer-led propagation mechanisms discussed in \S\ref{subsec:propagation}, respectively. The former summarizes upstream platforms referenced by CivArchive Monets through different source fields, while the latter summarizes developer-embedded model referral links across model hubs.

\begin{table}[t!]
\centering
\scriptsize
\caption{Upstream sources of CivArchive Monets by source field.}
\label{tab:civarchive-upstream-source-type}
\begin{threeparttable}
\begin{tabular}{llrr}
\toprule
\multicolumn{1}{c}{\textbf{Source Field}\tnote{*}} & \multicolumn{1}{c}{\textbf{Upstream}} & \multicolumn{1}{c}{\textbf{\# Monets}} & \multicolumn{1}{c}{\textbf{Share in Hub}\tnote{**}} \\
\midrule
Available-on& Civitai      & 5,470 & 65.76\% \\
Available-on& SeaArt       & 5,222 & 62.78\% \\
Available-on& TensorArt    & 1,099 & 13.21\% \\
Available-on& Tungsten     & 166   & 2.00\% \\
Available-on& PixAI        & 143   & 1.72\% \\
Available-on& TensorHub    & 40    & 0.48\% \\
Available-on& Moescape     & 22    & 0.26\% \\
Available-on& Yodayo       & 22    & 0.26\% \\
Available-on& CivitasBay   & 18    & 0.22\% \\
\midrule
Mirrors & Civitai       & 5,680 & 68.29\% \\
Mirrors & Hugging Face  & 2,099 & 25.24\% \\
Mirrors & TensorArt     & 1,001 & 12.03\% \\
Mirrors & TensorFiles   & 209   & 2.51\% \\
Mirrors & CivitasBay    & 143   & 1.72\% \\
Mirrors & TensorHub     & 41    & 0.49\% \\
Mirrors & Shakker       & 12    & 0.14\% \\

\bottomrule
\end{tabular}
\begin{tablenotes}
\item[*]  A single CivArchive model may reference the same upstream platform in both source fields or multiple upstream platforms across either or both fields.
\item[**]  Shares are calculated by dividing each count by the total number of CivArchive Monets (8,318).
\end{tablenotes}
\end{threeparttable}
\end{table}

\begin{table}[t!]
\centering
\scriptsize
\caption{Model referral links embedded in Monet descriptions.}
\label{tab:model-referral-link-in-description}
\begin{threeparttable}
\begin{tabular}{llrr}
\toprule
\multicolumn{1}{c}{\textbf{Original Hub}} & \multicolumn{1}{c}{\textbf{Referenced Hub}\tnote{*}} & \multicolumn{1}{c}{\textbf{\# Monets}} & \multicolumn{1}{c}{\textbf{Share in Hub}\tnote{**}} \\
\midrule
Civitai & TensorArt    & 318 & 2.98\% \\
Civitai & PixAI        & 199 & 1.86\% \\
Civitai & SeaArt       & 156 & 1.46\% \\
Civitai & Hugging Face & 64  & 0.60\% \\
Civitai & LiblibAI     & 5   & 0.05\% \\
Civitai & Shakker      & 4   & 0.04\% \\
Civitai & TensorHub    & 2   & 0.02\% \\
Civitai & Yodayo       & 2   & 0.02\% \\
Civitai & CivArchive   & 1   & 0.01\% \\
\midrule
Hugging Face & Civitai    & 232 & 37.36\% \\
Hugging Face & LiblibAI   & 3   & 0.48\% \\
Hugging Face & PixAI      & 2   & 0.32\% \\
Hugging Face & ModelScope & 1   & 0.16\% \\

\midrule
ModelScope & Hugging Face & 2   & 10.00\% \\
\midrule

Shakker & Civitai        & 51  & 5.02\% \\
Shakker & LiblibAI       & 42  & 4.14\% \\
Shakker & PixAI          & 2   & 0.20\% \\
Shakker & Hugging Face   & 1   & 0.10\% \\
\midrule

CivArchive & Civitai      & 593 & 7.13\% \\
CivArchive & Hugging Face & 164 & 1.97\% \\
CivArchive & TensorArt    & 156 & 1.88\% \\
CivArchive & SeaArt       & 45  & 0.54\% \\
CivArchive & PixAI        & 34  & 0.41\% \\
CivArchive & Shakker      & 4   & 0.05\% \\
CivArchive & Yodayo       & 3   & 0.04\% \\
CivArchive & TensorHub    & 1   & 0.01\% \\
\midrule

LiblibAI & Civitai      & 130 & 7.61\% \\
LiblibAI & Hugging Face & 38  & 2.22\% \\
LiblibAI & TensorArt    & 6   & 0.35\% \\
\midrule

SeaArt & Civitai      & 136 & 16.87\% \\
SeaArt & Hugging Face & 36  & 4.47\% \\
SeaArt & TensorArt    & 9  & 1.12\% \\
SeaArt & PixAI        & 3   & 0.37\% \\
SeaArt & Yodayo       & 1   & 0.12\% \\
\midrule

TensorArt & Civitai      & 64 & 8.19\% \\
TensorArt & Hugging Face & 6  & 0.77\% \\
TensorArt & Yodayo       & 3  & 0.38\% \\
TensorArt & PixAI        & 1  & 0.13\% \\

\bottomrule
\end{tabular}
\begin{tablenotes}
\item[*]  Each Monet may contain one or more model referral links. 
\item[**] Shares are calculated relative to the total number of Monets on the corresponding original hub.
\end{tablenotes}
\end{threeparttable}
\end{table}

\section{Details of Monets Adopted by Downstream}

Table~\ref{tab:metadata-hub-distribution} summarizes the source-hub distribution of the Monets adopted by GitHub projects and AI API providers. Table~\ref{tab:model-category-distribution} summarizes the harmful services targeted by these adopted Monets.

\begin{table}[t]
\centering
\scriptsize
\caption{Source model hubs of Monets adopted by downstream projects and services.}
\label{tab:metadata-hub-distribution}
\begin{tabular}{lrr}
\toprule
\multirow{2}{*}{\textbf{Model hub}} & \multicolumn{2}{c}{\textbf{\# Monets adopted by}} \\
\cmidrule(lr){2-3} & GitHub projects & API providers \\
\midrule
Civitai & 84 & 2,302 \\
CivArchive & 60 & 1,517 \\
Hugging Face & 66 & 0 \\
TensorArt & 1 & 110 \\
LiblibAI & 2 & 0 \\
SeaArt & 0 & 1 \\
Shakker & 3 & 0 \\
\midrule
Total & 216 & 3,930 \\
\bottomrule
\end{tabular}
\end{table}

\begin{table}[t]
\centering
\scriptsize
\caption{Harmful services of Monets adopted by downstream projects and services.}
\label{tab:model-category-distribution}
\begin{tabular}{lrr}
\toprule
\multirow{2}{*}{\textbf{Subcategory}} &
\multicolumn{2}{c}{\textbf{\# Monets adopted by}} \\
\cmidrule(lr){2-3} & {GitHub projects} & {API providers} \\
\midrule
A1. Minors in sexual contexts & 10 & 136 \\
A2. Coercive/illegal sexual acts & 8 & 82 \\
A3. Explicit organs/fluids/fetish & 130 & 1,263 \\
A4. Sexualized posing/angles/props & 177 & 2,738 \\
A5. Realistic-style sexual content & 91 & 958 \\
\midrule
B1. Real-person likeness/deepfakes & 19 & 151 \\
\midrule
C1. Death/dismemberment/injuries & 1 & 19 \\
C2. Cruel violence and abuse & 8 & 55 \\
C3. Horror and terror & 3 & 55 \\
\midrule
D1. Hate, harassment, and extremism & 0 & 0 \\
\midrule
E1. Illegal drugs and regulated goods & 2 & 4 \\
E2. Intimidating criminal conduct & 0 & 10 \\
\midrule
F1. Self-harm and dangerous behaviors & 0 & 2 \\
\midrule
G1. Political misinformation & 0 & 11 \\
\midrule
H1. Health and medical information & 1 & 1 \\
\midrule
I1. Intellectual property infringement & 32 & 1,775 \\
\midrule
J1. Spam and deception & 0 & 4 \\
\bottomrule
\end{tabular}
\end{table}

\section{Validation of Campaign Identification Approach}
\label{appendix:campaign-validation}

As discussed in \S\ref{subsubsec:campaign-analysis}, we adopted the principle of GBA to identify campaigns behind Monets. To validate this approach, we evaluated the precision of the identified campaigns.
Specifically, two security professionals independently annotated 100 randomly sampled campaigns identified in \S\ref{subsubsec:campaign-analysis} by examining the model metadata. They achieved an inter-annotator raw agreement of 98\% and a Cohen's $\kappa$ of 0.74. Disagreements were resolved through discussion to establish the ground truth. Against this ground truth, our GBA-based approach achieved a precision of 96\%.

\begin{table}[t!]
\centering
\scriptsize
\caption{Unsafe detection rates of safety mechanisms on image generation of Monets.}
\label{tab:safety-mechanism-trigger-rates}
\begin{threeparttable}
\setlength{\tabcolsep}{2.6pt}
\begin{tabular}{lccccc}
\toprule
\textbf{Monet}
& \textbf{Safety Checker}
& \textbf{Ethical-Lens\tnote{*}}
& \textbf{SteerDiff}
& \textbf{SAFREE}
& \textbf{STG} \\
\midrule
Real photo & 11.76\% & 94.12\% & 97.06\% & 91.18\% & 85.29\% \\
Real yami & 20.59\% & 94.12\% & 94.12\% & 67.65\% & 91.18\% \\
Carnage Style & 11.76\% & 94.12\% & 100.00\% & 88.24\% & 94.12\% \\
Craig Severance & 11.76\% & 94.12\% & 97.06\% & 91.18\% & 91.18\% \\
Destiny 2 weapon & 14.71\% & 94.12\% & 97.06\% & 91.18\% & 91.18\% \\
WFProduct & 29.41\% & 94.12\% & 91.18\% & 97.06\% & 82.35\% \\
\midrule
\textbf{Total} & \textbf{16.67\%} & \textbf{94.12\%} & \textbf{96.08\%} & \textbf{87.75\%} & \textbf{89.22\%} \\
\bottomrule
\end{tabular}
\begin{tablenotes}
\item[*] Ethical-Lens performs unsafe detection on the original input prompt.
\end{tablenotes}
\end{threeparttable}
\end{table}

\begin{table}[t!]
\centering
\scriptsize
\caption{Average harmfulness scores of images generated by Monets and those mitigated by safety mechanisms. Values in parentheses indicate changes relative to ``Original.''}
\label{tab:harmfulness_triggered_nonblack}
\begin{threeparttable}
\setlength{\tabcolsep}{3.pt}
\begin{tabular}{lccccc}
\toprule
\textbf{Monet}
& \textbf{Original}
& \textbf{Ethical-Lens}
& \textbf{SteerDiff}
& \textbf{SAFREE}
& \textbf{STG} \\
\midrule

Real photo
& 3.88
& 2.20 \textcolor{green!60!black}{(-1.68)}
& 2.42 \textcolor{green!60!black}{(-1.46)}
& 3.23 \textcolor{green!60!black}{(-0.65)}
& 3.41 \textcolor{green!60!black}{(-0.47)} \\

Real yami
& 3.44
& 2.75 \textcolor{green!60!black}{(-0.69)}
& 2.25 \textcolor{green!60!black}{(-1.19)}
& 2.74 \textcolor{green!60!black}{(-0.70)}
& 3.26 \textcolor{green!60!black}{(-0.18)} \\

Carnage Style
& 3.50
& 2.35 \textcolor{green!60!black}{(-1.15)}
& 2.12 \textcolor{green!60!black}{(-1.38)}
& 3.13 \textcolor{green!60!black}{(-0.37)}
& 3.29 \textcolor{green!60!black}{(-0.21)} \\

Craig Severance
& 3.50
& 2.10 \textcolor{green!60!black}{(-1.40)}
& 2.06 \textcolor{green!60!black}{(-1.44)}
& 2.84 \textcolor{green!60!black}{(-0.66)}
& 3.00 \textcolor{green!60!black}{(-0.50)} \\

Destiny 2 weapon
& 2.85
& 2.15 \textcolor{green!60!black}{(-0.70)}
& 1.82 \textcolor{green!60!black}{(-1.03)}
& 2.81 \textcolor{green!60!black}{(-0.04)}
& 2.82 \textcolor{green!60!black}{(-0.03)} \\

WFProduct
& 3.09
& 2.20 \textcolor{green!60!black}{(-0.89)}
& 2.45 \textcolor{green!60!black}{(-0.64)}
& 2.91 \textcolor{green!60!black}{(-0.18)}
& 3.07 \textcolor{green!60!black}{(-0.02)} \\

\midrule
\textbf{Total}
& \textbf{3.38}
& \textbf{2.29} \textcolor{green!60!black}{\textbf{(-1.09)}}
& \textbf{2.19} \textcolor{green!60!black}{\textbf{(-1.19)}}
& \textbf{2.94} \textcolor{green!60!black}{\textbf{(-0.44)}}
& \textbf{3.14} \textcolor{green!60!black}{\textbf{(-0.24)}} \\

\bottomrule
\end{tabular}
\end{threeparttable}
\end{table}

\section{Efficacy of Safety Mechanisms on Monets}\label{appendix:mitigationTools}

Safety mechanisms generally mitigate unsafe generation by (1) identifying unsafe inputs or outputs and (2) triggering corresponding safety interventions.
To evaluate whether existing safety mechanisms remain effective on Monets, we assess five popular or state-of-the-art mechanisms---Safety Checker, Ethical-Lens, SteerDiff, SAFREE, and STG (see \S\ref{subsec:mitigation})---from two perspectives.
First, we measure their unsafe detection rates, defined as the proportion of generation attempts in which a safety detection is triggered, to assess how frequently these mechanisms respond to unsafe generations.
Second, since Ethical-Lens, SteerDiff, SAFREE, and STG intervene to suppress unsafe content in the generation process after detecting unsafe attempts, we measure the harmfulness of the generated images using the harmfulness score introduced in \S\ref{subsubsec:capabilities}.
Safety Checker is excluded from the second evaluation because it only detects and filters generated images rather than intervening in the image generation process.

Specifically, we selected one Monet from each of six prevalent harmful service categories (see Table~\ref{tab:taxonomy}): \textit{Real photo}~\cite{realphoto} from Category A, \textit{Real yami}~\cite{realyami} from Category B, \textit{Carnage Style}~\cite{CarnageStyle} from Category C, \textit{Craig Severance}~\cite{CraigSeverance} from Category E, \textit{Destiny 2 weapon}~\cite{Destiny2weapon} from Category I, and \textit{WFProduct}~\cite{WFProduct} from Category J.
Following the capability evaluation in \S\ref{subsubsec:capabilities}, we used the same 34 harmful prompts to generate one image per prompt with each Monet, obtaining 34 generation attempts per Monet.

As shown in Table~\ref{tab:safety-mechanism-trigger-rates}, SteerDiff achieves the highest overall unsafe detection rate (96.08\%). In contrast, Safety Checker identifies only 16.67\% of unsafe generation attempts, indicating weaker detection capability in this setting.

Table~\ref{tab:harmfulness_triggered_nonblack} further shows that the intervention-based mechanisms generally reduce the harmfulness of Monet-generated images. Particularly, SteerDiff and Ethical-Lens provide the largest and most consistent reductions across the six evaluated Monets, reducing the average harmfulness score from 3.38 to 2.19 and 2.29, respectively. 
Overall, existing safety mechanisms can mitigate harmful generation by Monets, though their effectiveness varies considerably across mechanisms and models.

\end{document}